\documentclass[12pt]{article}
 
\usepackage{amssymb,cite}
\usepackage{epsfig}

\begin{document}

\begin{flushright}
CERN-TH/2003-096\\
BUTP-2003/09\\
April 2003\\
\end{flushright}

\begin{center}
\centerline{\large\bf An Analysis of the Inclusive Decay 
                      $\Upsilon (1S) \to \eta^\prime X$ and}
\smallskip 
\centerline{\large\bf Constraints on the $\eta^\prime$-Meson 
                      Distribution Amplitudes}

\vspace*{1.5cm}

{\large A.~Ali\footnote{On leave of absence from Deutsches 
        Elektronen-Synchrotron DESY, Hamburg, FRG.}
\vskip0.2cm
{\it Theory Division, CERN, CH-1211 Geneva 23, Switzerland}}\\
\vspace*{0.3cm}
\centerline{and}
\vspace*{0.3cm}
{\large A.Ya.~Parkhomenko\footnote{On leave of absence from 
        Department of Theoretical Physics,
        Yaroslavl State University,
        Sovietskaya~14, 150000 Yaroslavl, Russia.
}
\vskip0.2cm 
{\it Institut f$\ddot u$r Theoretische Physik, 
     Universit$\ddot a$t Bern, \\ 
     CH-3012 Bern, Switzerland }}

\vskip0.5cm
{\Large Abstract\\}
\vskip3truemm

\parbox[t]{\textwidth}{
We calculate the $\eta^\prime$-meson energy spectrum in the decay
$\Upsilon (1S) \to \eta^\prime g g g \to \eta^\prime X$ in the leading-order 
perturbative  QCD in the static quark  limit for the Orthoquarkonium. Our 
principal result is the extraction of parameters of the $\eta^\prime g^* g$ 
effective vertex function (EVF) involving a virtual and a real gluon from 
the available data on the hard part of the $\eta^\prime$-meson energy 
spectrum. The perturbative QCD based framework provides a good description 
of the available CLEO data, allowing to constrain the lowest Gegenbauer 
coefficients $B^{(q)}_2$ and $B^{(g)}_2$ of the quark-antiquark and 
gluonic distribution amplitudes of the $\eta^\prime$-meson. The resulting 
constraints are combined with the existing ones on these 
coefficients from an analysis of the $\eta-\gamma$ and $\eta^\prime-\gamma$
transition form factors and the requirement of positivity of the EVF, 
yielding $B^{(q)}_2(\mu_0^2) = -0.008 \pm 0.054$ and $B^{(g)}_2(\mu_0^2) = 
4.6 \pm 2.5$ for $\mu_0^2 = 2$~GeV$^2$. This reduces significantly the 
current uncertainty on these coefficients. The resulting EFV 
$F_{\eta^\prime g^* g} (p^2, 0, m_{\eta^\prime}^2)$, including the 
$\eta^\prime$-meson mass effects, is presented.
}

\end{center}

\thispagestyle{empty}
\newpage
\setcounter{page}{1}
\textheight 23.0 true cm
\section{Introduction} 
\label{sec:introduction} 
Experiments involving the production and decays of the $\eta$- and 
$\eta^\prime$-mesons
are consistent with the picture that the $\eta$-meson is largely 
an $SU(3)_{\rm F}$ flavour-octet state, but the $\eta^\prime$-meson 
contains a significant amount of a $U(3)_{\rm F}$ flavour-singlet 
quark-antiquark ($\bar q q$) component in its wave-function
\cite{Gilman:1987ax,Leutwyler:1997yr,Kaiser:2000gs,Feldmann:1998vh,Feldmann:1998sh}; 
in addition, the gluonic ($gg$) content of the $\eta^\prime$-meson is 
substantial~\cite{Ball:1995zv}. This implies that for  
processes involving gluons and the $\eta^\prime$-meson,
in particular in the decays of heavy 
mesons of current phenomenological interest, such as
 $b \to s g \eta^\prime$ inducing the $B \to \eta^\prime X_s$ decay 
\cite{Atwood:1997bn,Hou:1997wy,Kagan:1997qn,Halperin:1997ma,Yuan:1997ts,Datta:1997nr,Ahmady:1997fa}
and the exclusive $B \to (\eta,\eta^\prime)(K,K^*)$ decays 
\cite{Halperin:1997as,Cheng:1997if,Dighe:1997hm,Deshpande:1997ar,Du:1997hs,Ali:1997ex,Ali:1998eb,Beneke:2002jn},
but also the inclusive decay of the Orthoquarkonium  
$\Upsilon(1S) \to gg(g^* \to \eta^\prime g) \to  \eta^\prime 
X$~\cite{Kagan:2002dq}, the effective $\eta^\prime$-gluon-gluon vertex 
plays an important role. This effective vertex function (EFV) has to be known 
sufficiently well to undertake a quantitative 
analysis of the data involving the $\eta^\prime$-meson. Calling this EFV 
 $F_{\eta^\prime g^* g^*}(p_1^2, p_2^2, m_{\eta^\prime}^2)$,
where $p_1^2$ and $p_2^2$ are the virtualities of the two gluons, the QCD 
anomaly provides the normalization of this vertex for on-shell gluons, 
$F_{\eta^\prime g  g}(0,0,m_{\eta^\prime}^2)$.
When one or both of the gluons are virtual   
with relatively large virtuality, the effective
$\eta^\prime g^* g^{(*)}$ vertex can be calculated in
perturbative QCD~\cite{Muta:1999tc,Ali:2000ci,Kroll:2002nt,Agaev:2002ek}.

In this approach, the $\eta^\prime$-meson wave-function is
described in terms of two light-cone distribution amplitudes (LCDAs) 
involving the quark-antiquark~$\phi^{(q)}_{\eta^\prime} (x, Q^2)$ and
the gluonic~$\phi^{(g)}_{\eta^\prime} (x, Q^2)$ components, where~$x$ 
is the scaled 
energy of one of the partons of the $\eta^\prime$-meson and~$Q^2$ is   
the typical hard scale in the vertex. These two components  mix if the QCD
evolution is taken into account. The leading-twist LCDAs of the 
$\eta^\prime$-meson
can be expressed as  infinite series of the Gegenbauer polynomials
$C^{3/2}_n (x - \bar x)$ (for the quark-antiquark) and
$C^{5/2}_{n -1} (x - \bar x)$ (for the gluonic) components
\cite{Terentev:qu,Ohrndorf:1981uz,Shifman:1981dk,Baier:1981pm,Terentev:wv,Belitsky:1998gc}
\begin{equation}
\phi^{(q)}_{\eta^\prime} (x, Q^2) = 6 x \bar x
\left [ 1 +
\sum_{{\rm even} \, n \ge 2} A_n (Q^2) \, C^{3/2}_n (x - \bar x)
\right ] ,
\label{eq:DAq-gen}
\end{equation}
\begin{equation}
\phi^{(g)}_{\eta^\prime} (x, Q^2) = x^2 \bar x^2
\sum_{{\rm even} \, n \ge 2}
B_n (Q^2) \, C^{5/2}_{n - 1} (x - \bar x) ,
\label{eq:DAg-gen}
\end{equation}
where $\bar x = 1 - x$, and the following notation is used
for the Gegenbauer moments:
\begin{eqnarray}
A_n (Q^2) & = & B^{(q)}_n (\mu_0^2) \left (
\frac{\alpha_s (\mu_0^2)}{\alpha_s (Q^2)} \right )^{\gamma_+^n} 
+ \rho_n^{(g)} \, B^{(g)}_n (\mu_0^2) \left (
\frac{\alpha_s (\mu_0^2)}{\alpha_s (Q^2)} \right )^{\gamma_-^n} ,
\label{eq:An} \\
B_n (Q^2) &= & \rho_n^{(q)} \, B^{(q)}_n (\mu_0^2) \left (
\frac{\alpha_s (\mu_0^2)}{\alpha_s (Q^2)} \right )^{\gamma_+^n}
+ B^{(g)}_n (\mu_0^2) \left (
\frac{\alpha_s (\mu_0^2)}{\alpha_s (Q^2)} \right )^{\gamma_-^n} .
\label{eq:Bn}
\end{eqnarray}
The constrained parameters~$\rho^{(q)}_n$, $\rho^{(g)}_n$, $\gamma_+^n$
and~$\gamma_-^n$ are computable and can be found, for example, in the 
Appendix~A of 
Ref.~\cite{Ali:2000ci}. Usually, one employs an approximate form for the
$\eta^\prime$-meson LCDAs in which only the first
non-asymptotic terms in both the quark-antiquark and gluonic components 
are kept. Thus, in this approximation
\begin{equation}
\phi^{(q)}_{\eta^\prime} (x, Q^2) = 6 x \bar x
\left [ 1 + 6 (1 - 5 x \bar x) \, A_2 (Q^2) \right ] ,
\qquad
\phi^{(g)}_{\eta^\prime} (x, Q^2) =
5 x^2 \bar x^2 \, (x - \bar x) \, B_2 (Q^2) ,
\label{eq:DAqg}
\end{equation}
where the explicit forms for $C^{3/2}_2 (x - \bar x)$ and
$C^{5/2}_1 (x - \bar x)$ have been used.
The free parameters~$B^{(q)}_2 (\mu_0^2)$ and~$B^{(g)}_2 (\mu_0^2)$, 
which enter in $A_2(Q^2)$ and $B_2(Q^2)$, are not determined
from first principles, and have to be modeled or extracted from a
phenomenological analysis of  
data. To that end, a fit of the CLEO and L3 data on the  
$\eta - \gamma$ and $\eta^\prime - \gamma$ transition form factors 
for~$Q^2$ larger
than 2~GeV$^2$~\cite{Gronberg:1997fj,Acciarri:1997yx} was recently 
undertaken in Ref.~\cite{Kroll:2002nt}, yielding
\begin{equation}
A_2 (1~{\rm GeV}^2) = - 0.08 \pm 0.04,
\qquad
B_2 (1~{\rm GeV}^2) = 9 \pm 12,
\label{eq:AB2-fit}
\end{equation}
where the initial scale of the evolution is taken as
$\mu_0^2 = 1$~GeV$^2$. Note, that the coefficients  
$A_2 (1~{\rm GeV}^2)$ and~$B_2 (1~{\rm GeV}^2)$ are strongly correlated. 
The estimates~(\ref{eq:AB2-fit})
can be translated in terms of the universal free
parameters~$B^{(q)}_2$ and~$B^{(g)}_2$,
 yielding:
\begin{equation}
B^{(q)}_2(1~{\rm GeV}^2) = 0.02 \pm 0.17,
\qquad
B^{(g)}_2(1~{\rm GeV}^2) = 9.0 \pm 11.5 .
\label{eq:B2-fit}
\end{equation}
The current determination of these coefficients, in particular 
$B^{(g)}_2$, is rather poor, leading to a huge uncertainty in the 
evaluation of the $\eta^\prime g^* g^{(*)}$ vertex function 
$F_{\eta^\prime g^* g^{(*)}} (p_1^2, p_2^2, m_{\eta^\prime}^2)$.  

In this paper, we undertake a perturbative-QCD based analysis of
the recent data on the inclusive process $\Upsilon (1S) \to \eta^\prime X$ 
in the large $\eta^\prime$-meson energy region published recently by the 
CLEO collaboration \cite{Artuso:2002px}, which is expected to be dominated 
by the process $\Upsilon (1S) \to \eta^\prime g g g \to \eta^\prime X$. 
Moreover, Chen and Kagan~\cite{Kagan:2002dq} have argued that the
shape of the $\eta^\prime$-meson energy spectrum in this decay is 
sensitive to the shape of the 
$F_{\eta^\prime g^* g} (p^2, 0, m_{\eta^\prime}^2)$ vertex function, 
involving the $\eta^\prime$-meson, a real and a virtual gluons (see  
Fig.~\ref{fig:diagram}). This sensitivity has already been used by 
the CLEO collaboration~\cite{Artuso:2002px} to rule out
certain models for this vertex, with the CLEO analysis favouring a
rapidly falling $p^2$-dependence of the vertex $F_{\eta^\prime g^* g} 
(p^2, 0, m_{\eta^\prime}^2)$, in line with the perturbative
QCD predictions~\cite{Muta:1999tc,Ali:2000ci}. Motivated by these 
observations, we
undertake a quantitative analysis of the CLEO data to constrain the 
LCDAs involving the quark-antiquark and the gluonic components of the
$\eta^\prime$-meson. The results of this analysis are presented in terms 
of the coefficients $B^{(q)}_2 (2~{\rm GeV}^2)$ 
and $B^{(g)}_2 (2~{\rm GeV}^2)$,
where we have taken the initial scale as 
$\mu_0^2 = m_{\eta^\prime}^2 + p_0^2 = 2~{\rm GeV}^2$, 
which corresponds to the minimum gluon virtuality 
$p_0^2 \simeq 1~{\rm GeV}^2$.
This analysis is then combined with an earlier analysis of  
the $\eta - \gamma$ and $\eta^\prime - \gamma$ transition form 
factors~\cite{Kroll:2002nt} to further constrain the two parameters.
As the physical interpretation of the function $F_{\eta^\prime g^* g} 
(p^2, 0, m_{\eta^\prime}^2)$ is that it represents a probability 
distribution, much the same way as the partonic distributions are in,
for example, deep inelastic lepton scattering off nucleons, this function
must remain positive definite in the entire~$p^2$ range. The requirement
of positive definiteness of the function $F_{\eta^\prime g^*g} 
(p^2, 0, m_{\eta^\prime}^2)$ provides additional constraints on the 
parameters $B^{(q)}_2 (2~{\rm GeV}^2)$ and $B^{(g)}_2 (2~{\rm GeV}^2)$,
in particular on the latter.  
The combined analysis leads to the following correlated ranges for 
these coefficients: 
\begin{equation}
B^{(q)}_2 (2~{\rm GeV}^2) = - 0.008 \pm 0.054~,
\qquad
B^{(g)}_2 (2~{\rm GeV}^2) = 4.6 \pm 2.5~.
\label{eq:B2-fitfinal}
\end{equation}
Finally, we use this information  to calculate     
the $F_{\eta^\prime g^* g} (p^2,0,m_{\eta^\prime}^2)$ vertex, 
including the $\eta^\prime$-meson mass effects,  relegating 
the detailed derivation to a subsequent paper~\cite{Ali:2003}.

This paper is organized as follows: In section~2, we calculate the 
branching ratio for the process $\Upsilon (1S) \to ggg^* (g^* \to 
\eta^\prime g) \to \eta^\prime X$ and the $\eta^\prime$-meson energy 
spectrum in this decay. Numerical analysis of the CLEO data is carried 
out in section~3, and the resulting constraints are combined with the 
analysis of the $\eta^\prime - \gamma$ transition form factor to
determine the LCDAs of the $\eta^\prime$-meson. 
Section~4 contains a brief summary. 
The expressions for the matrix element squared for the decay
$\Upsilon(1S) \to ggg^* (g^* \to \eta^\prime g)$ are displayed 
in the Appendix.

\section{Branching Ratio and $E_{\eta^\prime}$-Distribution in the 
         Decay $\Upsilon (1S) \to \eta^\prime g g g \to \eta^\prime X$}
\label{sec:Branch}

Several processes  contribute to the inclusive $\eta^\prime$-meson 
production in the $\Upsilon (1S)$-meson decay. The two dominant decays 
are: $\Upsilon (1S) \to \gamma^* \to q \bar q \to \eta^\prime X$ and
$\Upsilon (1S) \to g g g^* (g^* \to \eta^\prime g) \to \eta^\prime X$.
The first of these has been estimated from the measured value of the
hadronic cross-section $R_{\rm had}$ and the branching ratio ${\cal 
B}(\Upsilon(1S) \to \mu^+ \mu^-)$, yielding ${\cal B}(\Upsilon (1S) \to 
\gamma^*  \to q \bar q) =(8.8 \pm 0.3)\%$~\cite{Artuso:2002px}, and the
CLEO data has been corrected for the $q\bar q$ component. In 
addition, there is also the continuum production from the 
process $e^+ e^- \to \gamma^* \to q \bar q \to \eta^\prime X$, which can 
be estimated  from the $e^+ e^-$-continuum data below the resonance. This
data also provides a good profile of the  
fragmentation in the process $\gamma^* \to q \bar q \to \eta^\prime X$.
Typically, the fragmentation processes 
involving the intermediate $q \bar q$ state populate the low-$z$ region, 
where $z \equiv E_{\eta^\prime}/E_{\rm beam} = 2 E_{\eta^\prime}/M$ is 
the relative $\eta^\prime$-meson energy expressed in terms  
of the $\Upsilon(1S)$-meson mass~$M$. 
One expects that in the large-$z$ region, the process $\Upsilon(1S) \to 
ggg^* (g^* \to g \eta^\prime)$ dominates. Assuming this, we will 
concentrate here on the intermediate three-gluon $(ggg^*)$ state and 
analyze the $E_{\eta^\prime}$-spectrum in the large-$z$ region alone. 
The quality of the fits provides a justification of this
procedure {\it a posteriori}. 

\begin{figure}[tb]
\centerline{
\psfig{file=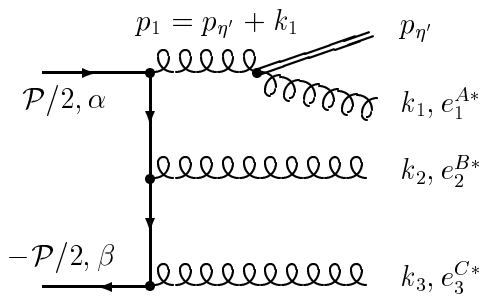,width=.45\textwidth}
}
\caption{\label{fig:diagram}%
     A typical Feynman diagram describing the $\Upsilon (1S) \to 
     g g g^* (g^* \to \eta^\prime g) \to \eta^\prime X$ decay. 
     The directed solid line denotes the quark~$b$ and 
     antiquark~$\bar b$ of the $\Upsilon$-meson and the double 
     solid line denotes the $\eta^\prime$-meson.}    
\end{figure}

A typical Feynman diagram describing the decay $\Upsilon (1S) \to 
g g g^* (g^* \to \eta^\prime g) \to \eta^\prime X$ is presented in 
Fig.~\ref{fig:diagram}. 
%
%
%
%
The other 17~diagrams can be obtained from the above one by 
the permutations  of the gluons in the intermediate (virtual) 
and final state. For the matrix element calculations, 
we adopted the static limit for the heavy quark-antiquark pair in 
the orthoquarkonium $\Upsilon (1S)$, so that both the quark~($b$) 
and antiquark~($\bar b)$ carry half of the $\Upsilon (1S)$-meson 
four-momentum, and the velocity-dependent corrections are neglected.  
 
The total decay amplitude can be divided into three parts: 
\begin{equation}
{\cal M} [\Upsilon ({\cal P}) \to \eta^\prime (p_{\eta^\prime})
          g (k_1) g (k_2) g (k_3)] = 
\sum_{i = 1}^3 {\cal M}_i , 
\label{eq:ME-Up-3gep-total} 
\end{equation}
where each of the three terms ${\cal M}_i$ collects the contributions 
from the diagrams with the virtual gluon of the same four-momentum 
$p_i = p_{\eta^\prime} + k_i$, with~$p_{\eta^\prime}$ and~$k_i$ being 
the four-momenta of the $\eta^\prime$-meson and the $i$th final gluon. 
The explicit forms of the three contributions~${\cal M}_i$  
are as follows: 
\begin{eqnarray}
{\cal M}_1 & = & \frac{d_{A B C}}{4 \sqrt{N_c}} \,
\frac{g_s^3 \, \sqrt M \, \psi (0) \, F_{\eta^\prime g} (p_1^2)}
     {({\cal P} k_2) ({\cal P} k_3) (p_1 p_1^\prime)} \,
\bigg \{ \frac{4}{M^2} \, ( p_1 \tilde f_1^{A*}
[ f_2^{B*} f_3^{C*} + \tilde f_2^{B*} \tilde f_3^{C*} ]
f_V {\cal P} )
\label{eq:ME1} \\  
& + &
(f_2^{B*} \tilde f_3^{C*}) (f_1^{A*} f_V) +
(f_V \tilde f_3^{C*}) (f_1^{A*} f_2^{B*}) +
(f_V \tilde f_2^{B*}) (f_1^{A*} f_3^{C*})
\nonumber \\
& - & \frac{2}{p_1^2} \left [
(f_2^{B*} f_3^{C*}) (p_1 \tilde f_1^{A*} f_V p_1) +
(f_V f_3^{C*}) (p_1 \tilde f_1^{A*} f_2^{B*} p_1) +
(f_V f_2^{B*}) (p_1 \tilde f_1^{A*} f_3^{C*} p_1)
\right.
\nonumber \\
& + &
\left.
(f_2^{B*} \tilde f_3^{C*}) (p_1 f_1^{C*} f_V p_1) +
(f_V \tilde f_3^{C*}) (p_1 f_1^{A*} f_2^{B*} p_1) +
(f_V \tilde f_2^{B*}) (p_1 f_1^{A*} f_3^{C*} p_1)
\right ] \bigg \} , 
\nonumber
\end{eqnarray}
\begin{eqnarray}
{\cal M}_2 
& = & \frac{d_{A B C}}{4 \sqrt{N_c}} \,
\frac{g_s^3 \, \sqrt M \, \psi (0) \, F_{\eta^\prime g} (p_2^2)}
     {({\cal P} k_1) ({\cal P} k_3) (p_2 p_2^\prime)} \,
\bigg \{ \frac{4}{M^2} \, ( p_2 \tilde f_2^{B*}
[ f_3^{C*} f_1^{A*} + \tilde f_3^{C*} \tilde f_1^{A*} ]
f_V {\cal P} )
\label{eq:ME2} \\  
& + &
(f_1^{A*} \tilde f_3^{C*}) (f_2^{B*} f_V) +
(f_V \tilde f_1^{A*}) (f_2^{B*} f_3^{C*}) +
(f_V \tilde f_3^{C*}) (f_1^{A*} f_2^{B*})
\nonumber \\
& - & \frac{2}{p_2^2} \left [
(f_1^{A*} f_3^{C*}) (p_2 \tilde f_2^{B*} f_V p_2) +
(f_V f_1^{A*}) (p_2 \tilde f_2^{B*} f_3^{C*} p_2) +
(f_V f_3^{C*}) (p_2 \tilde f_2^{B*} f_1^{A*} p_2)
\right.
\nonumber \\
& + &
\left.
(f_1^{A*} \tilde f_3^{C*}) (p_2 f_2^{B*} f_V p_2) +
(f_V \tilde f_1^{A*}) (p_2 f_2^{B*} f_3^{C*} p_2) +
(f_V \tilde f_3^{C*}) (p_2 f_2^{B*} f_1^{A*} p_2)
\right ] \bigg \} , 
\nonumber
\end{eqnarray}
\begin{eqnarray}
{\cal M}_3 & = & \frac{d_{A B C}}{4 \sqrt{N_c}} \,
\frac{g_s^3 \, \sqrt M \, \psi (0) \, F_{\eta^\prime g} (p_3^2)}
     {({\cal P} k_1) ({\cal P} k_2) (p_3 p_3^\prime)} \,
\bigg \{ \frac{4}{M^2} \, ( p_3 \tilde f_3^{C*}
[ f_1^{A*} f_2^{B*} + \tilde f_1^{A*} \tilde f_2^{B*} ]
f_V {\cal P} )
\label{eq:ME3} \\  
& + &
(f_1^{A*} \tilde f_2^{B*}) (f_3^{C*} f_V) +
(f_V \tilde f_2^{B*}) (f_1^{A*} f_3^{C*}) +
(f_V \tilde f_1^{A*}) (f_2^{B*} f_3^{C*})
\nonumber \\
& - & \frac{2}{p_3^2} \left [
(f_1^{A*} f_2^{B*}) (p_3 \tilde f_3^{C*} f_V p_3) +
(f_V f_2^{B*}) (p_3 \tilde f_3^{C*} f_1^{A*} p_3) +
(f_V f_1^{A*}) (p_3 \tilde f_3^{C*} f_2^{B*} p_3)
\right.
\nonumber \\
& + &
\left.
(f_1^{A*} \tilde f_2^{B*}) (p_3 f_3^{C*} f_V p_3) +
(f_V \tilde f_2^{B*}) (p_3 f_3^{C*} f_1^{A*} p_3) +
(f_V \tilde f_1^{A*}) (p_3 f_3^{C*} f_2^{B*} p_3)
\right ] \bigg \} , 
\nonumber
\end{eqnarray}
where $d_{A B C}$ ($A, B, C = 1, \ldots, N_c^2 - 1$ are the colours
of the gluons) is the symmetrical constant of the colour $SU (N_c)$
group with $N_c = 3$; $g_s$ is the strong coupling constant; 
$(f_V)_{\mu \nu} = {\cal P}_\mu \eta_\nu - {\cal P}_\nu \eta_\mu$ 
is the polarization tensor of the $\Upsilon (1S)$-meson,  
$M$, ${\cal P}_\mu$, $\eta_\mu$ and $\psi (0)$ are the mass of the 
quarkonium, the four-momentum, the polarization vector, and 
the non-relativistic wave-function in the position space evaluated 
at the origin, respectively;   
$(f_i^A)_{\mu \nu} = k_{i \mu} e_{i \nu}^A - k_{i \nu} e_{i \mu}^A$ 
and $(\tilde f_i^A)_{\mu \nu} = \varepsilon_{\mu \nu \rho \sigma}
k_i^\rho e_i^{A \sigma}$ ($i = 1, 2, 3$) are the gluonic field-strength 
tensor and its dual involving the $i$th gluon with the polarization 
vector~$e_{i\mu}^A$ and four-momenta~$k_{i\mu}$;  
$p_i = p_{\eta^\prime} + k_i$ is the four-momentum of the virtual 
gluon, and $p_i^\prime = {\cal P} - p_i$. In the above equations 
the notations are as follows: 
$(f_1^A f_2^B) = (f_1^A)_{\mu \nu} (f_2^B)^{\nu \mu}$, 
$(p_1 f_1^A f_2^B p_1) = p_1^\mu (f_1^A)_{\mu \nu} (f_2^B)^{\nu \rho} 
p_{1\rho}$, etc. It is worth noting that the matrix element satisfies 
the Bose symmetry under the exchange of gluons, in particular, under 
the interchange of the first and second gluons  
${\cal M}_1 \leftrightarrow {\cal M}_2$ while the term~${\cal M}_3$ 
remains unchanged.  

The function $F_{\eta^\prime g} (p_i^2)$ is the effective 
vertex function involving the $\eta^\prime$-meson and 
two gluons, one of which is on the mass shell ($k_i^2 = 0$). We shall call
this interchangeably also the $\eta^\prime - g$ transition form factor. 
A form of this 
function motivated by the QCD analysis of the $\eta^\prime g^* g$ loop 
diagram was suggested by Kagan and Petrov~\cite{Kagan:1997qn}:  
\begin{equation}
F_{\eta^\prime g} (p^2) \equiv 
F_{\eta^\prime g^* g} (p^2, 0, m_{\eta^\prime}^2) =
\frac{m_{\eta^\prime}^2 \, H (p^2)}{p^2 - m_{\eta^\prime}^2} ,
\label{eq:VF}
\end{equation}
where the function~$H (p^2)$ was approximated by the constant
value $H (p^2) \approx H_0 \simeq 1.7 \, {\rm GeV}^{-1}$, 
extracted from the $J/\psi \to \eta^\prime \gamma$ 
decay~\cite{Atwood:1997bn}. In a companion paper~\cite{Ali:2003} 
we argue that the form~(\ref{eq:VF}) for the $\eta^\prime - g$ 
transition form factor also emerges in the perturbative calculations 
of this function in the hard-scattering approach by keeping the 
$\eta^\prime$-meson mass. In this approach, the dependence of 
the function~$H (p^2)$ on the gluon virtuality~$p^2$ is given 
by the following expression: 
\begin{equation}
H (p^2) = \frac{4 \pi \alpha_s (Q^2)}{m_{\eta^\prime}^2} \,
\sqrt 3 f_{\eta^\prime} \left [ 1 + A_2 (Q^2) - \frac{5}{6} \,
B_2 (Q^2) \, G_2^{(g)} (1, \zeta) \right ] ,
\label{eq:H-HSA}
\end{equation}
where $f_{\eta^\prime} \simeq 2 f_\pi/\sqrt 3$ is the
$\eta^\prime$-meson decay constant expressed in terms of
the $\pi$-meson decay constant $f_\pi \simeq 133$~MeV,
$\zeta = m_{\eta^\prime}^2/p^2$, and the function
$G_2^{(g)} (1, \zeta)$ has the form~\cite{Ali:2003}:
\begin{equation}
G_2^{(g)} (1, \zeta) = \frac{5}{3 \zeta} + \frac{2}{\zeta^2}
- \frac{4}{\zeta^3} - \frac{1}{\zeta}
\left [ 1 -  \frac{1}{\zeta} \right ]
\left [ 1 -  \frac{4}{\zeta^2} \right ] \ln (1 - \zeta) .
\label{eq:G2g}
\end{equation}
In Eq.~(\ref{eq:H-HSA}) the scale~$Q^2$ in the strong
coupling~$\alpha_s (Q^2)$, and also in the second Gegenbauer
moments~$A_2 (Q^2)$ and~$B_2 (Q^2)$ of the quark-antiquark 
and gluonic LCDAs of the $\eta^\prime$-meson, is related with 
the gluon virtuality~$p^2$, but there is an uncertainty in its
precise definition. One of the possibilities is  
to require that the function~$H (p^2)$ is finite at all values
of~$p^2$ including the singularity point of the EVF~(\ref{eq:VF}),   
$p^2 = m_{\eta^\prime}^2$, which can be done, for example, by
putting $Q^2 = |p^2| + m_{\eta^\prime}^2$.

\begin{figure}[tb]
\centerline{
\psfig{file=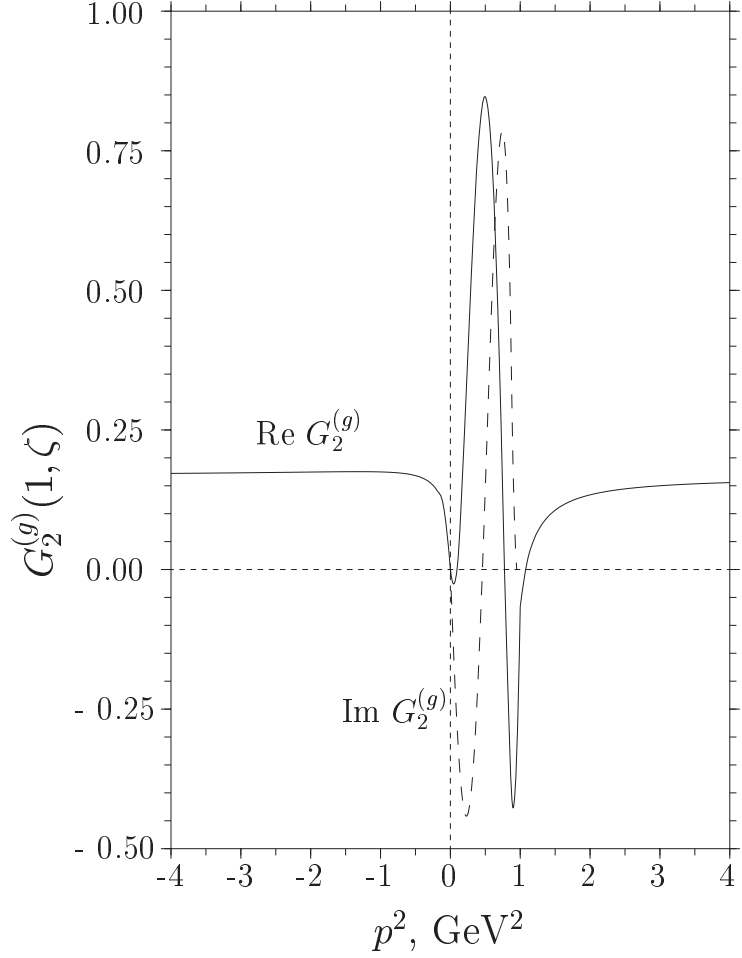,width=.65\textwidth}
}
\caption{\label{fig:Gg2}%
         The real (solid curve) and imaginary (dashed curve)   
         parts of the function~$G_2^{(g)} (1, \zeta)$, where
         $\zeta = m_{\eta^\prime}^2/p^2$, as a function of the
         momentum squared~$p^2$ of the virtual gluon. We use this 
         function only in the $|p^2| > 1~{\rm GeV}^2$ region.}
\end{figure}

The dependence of the function~$G_2^{(g)}$ on the momentum squared~$p^2$
of the virtual gluon is presented in Fig.~\ref{fig:Gg2} with the
value $G_2^{(g)} (1, 0) = 1/6$ in the large-$|p^2|$ asymptotic region. 
Since this result is based on the application of perturbation theory, 
its validity is restricted to the region $|p^2| > p_0^2$, where typically
$p_0^2 = 1~{\rm GeV}^2$ (or somewhat higher). In view of this, 
we shall set ${\rm Im}~G_2^{(g)} = 0$ and ignore the structure in 
${\rm Re}~G_2^{(g)}$ in the low-$\vert p^2 \vert$ region seen in 
Fig.~\ref{fig:Gg2}. 
 %
%
%
%

As already noted, the functions~$A_2 (Q^2)$ and~$B_2 (Q^2)$ 
in Eq.~(\ref{eq:H-HSA}) 
are the second Gegenbauer moments of the quark-antiquark 
and gluonic light-cone distribution amplitudes (LCDAs) of the 
$\eta^\prime$-meson. The scale~$\mu_0$ appearing in the definitions 
of these moments is set by the initial value in the evolution of 
the Gegenbauer moments.
As the mass of the $\eta^\prime$-meson is large, $\sim 1$~GeV,  
it is not a good approximation to neglect it. Assuming further, 
that the evolution of the Gegenbauer moments is valid for
the gluons with virtualities larger than 1~GeV$^2$, we shall take
$\mu_0^2 = Q_0^2 \simeq 2$~GeV$^2$.

The total decay width of the $\Upsilon (1S)$-meson into the 
$\eta^\prime$-meson and three gluons, $\Upsilon \to \eta^\prime g g g$,
averaged over the quarkonium spin states and summed over the
polarizations and colours of the final gluons can be written
in the form:
\begin{eqnarray}
\Gamma_{\eta^\prime X} \equiv
\Gamma [\Upsilon \to \eta^\prime g g g] & = & \frac{1}{3!} \,
\frac{1}{(2 \pi)^8} \, \frac{1}{2 M}
\int \frac{d{\bf k}_1}{2 \omega_1} \, \frac{d{\bf k}_2}{2 \omega_2} \,
\frac{d{\bf k}_3}{2 \omega_3} \,  
\frac{d{\bf p_{\eta^\prime}}}{2 E_{\eta^\prime}} 
\label{eq:DW-total} \\
& \times &
\delta^{(4)} ({\cal P} - k_1 - k_2 - k_3 - p_{\eta^\prime}) \, 
\frac{1}{3} \sum
\left |{\cal M} [\Upsilon \to \eta^\prime g g g] \right |^2 ,
\nonumber
\end{eqnarray}
where the factor $1/3!$ takes into account the identity of the
final gluons. The expression for the matrix element squared is 
rather lengthy and can be found in the Appendix, where we have also 
discussed some technical details of our Monte Carlo integration. 

The $\eta^\prime$-meson energy distribution function can be 
defined as follows: 
\begin{eqnarray}
\frac{dn}{dz} & = & \frac{1}{\Gamma_{3g}^{(0)}} \,
\frac{d \Gamma_{\eta^\prime X} (z)}{dz} =
\frac{1}{\Gamma_{3g}^{(0)}} \, \frac{1}{3!} \,
\frac{1}{(2 \pi)^8} \, \frac{1}{2 M}
\int \frac{d{\bf k}_1}{2 \omega_1} \, \frac{d{\bf k}_2}{2 \omega_2} \,
\frac{d{\bf k}_3}{2 \omega_3} \,  
\frac{d{\bf p_{\eta^\prime}}}{2 E_{\eta^\prime}}   
\label{eq:BR-spectrum} \\ 
& \times &
\delta^{(4)} ({\cal P} - k_1 - k_2 - k_3 - p_{\eta^\prime}) \, 
\delta (z - 2 E_{\eta^\prime}/M) \,  \frac{1}{3} \sum
\left |{\cal M} [\Upsilon \to \eta^\prime g g g] \right |^2 ,
\nonumber
\end{eqnarray}
where $\Gamma_{3g}^{(0)}$ is the three-gluon decay width of the 
$\Upsilon (1S)$-meson in the leading order: 
\begin{equation}
\Gamma_{3 g}^{(0)} =
\frac{16}{9} \, \left ( \pi^2 - 9 \right ) C_F \, B_F \,
\alpha_s^3 (M^2) \, \frac{|\psi (0)|^2}{M^2} .   
\label{eq:Gamma-3g}
\end{equation}
Here, $C_F = (N_c^2 - 1)/(2 N_c)$ and $B_F = (N_c^2 - 4)/(2 N_c)$
are the constants of the colour $SU (N_c)$ group, and 
$\alpha_s (M^2) = g_s^2 (M^2)/(4 \pi)$ is the strong coupling   
estimated at the scale of the $\Upsilon$-meson mass. The $\alpha_s$ 
corrections to the decay width $\Gamma_{3 g}$ are known since a long time 
and are numerically large~\cite{Mackenzie:1981sf}. However, we do not 
take them into account, as we have calculated the decay $\Upsilon(1S) 
\to ggg^*(g^* \to \eta^\prime g)$ only in the leading order.
One anticipates that in the distribution~$dn/dz$, a good part of the 
explicit $O(\alpha_s)$ corrections should drop out, and we are tacitly 
assuming that the remaining  corrections do not greatly influence the 
energy  spectrum derived in the lowest non-trivial order.   

\section{Numerical Analysis of the $\Upsilon(1S) \to \eta^\prime X$ Data 
and the $\eta^\prime$-Meson LC Distribution Amplitudes}
\label{sec:numeric}

As the results for the matrix element squared in the decay 
$\Upsilon(1S) \to ggg^* (g^* \to \eta^\prime g)$ are not yet available 
in the literature, we shall give their explicit expressions in this paper.
In that context we note that it is sufficient to have the expressions
for one of the diagonal terms $|{\cal M}_i|^2$ ($i = 1, 2, 3$) and one
of the non-diagonal terms ${\cal M}_i {\cal M}_j^*$ ($i \neq j$), as the
others can be obtained by using the Bose symmetry. With this, the
expressions for the components $(1/3) \sum |{\cal M}_1|^2$, derived 
from Eq.~$(\ref{eq:ME1})$, and $(1/3) \sum {\cal M}_1 {\cal M}_2^* + $~c.c., 
resulting from the cross terms in Eq.~$(\ref{eq:ME1})$ and~$(\ref{eq:ME2})$ 
in $\vert {\cal M}\vert^2$, are given in the Appendix.
 
\begin{figure}[tb]
\centerline{
\psfig{file=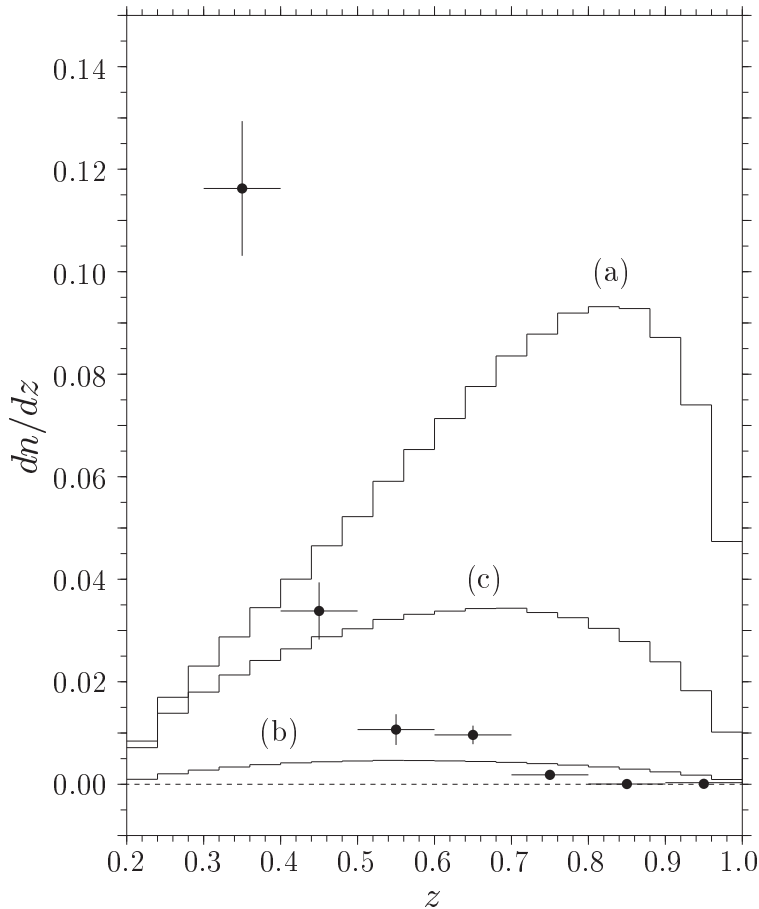,width=.65\textwidth}
}
\caption{\label{fig:CLEO1}%
         The energy spectrum of the $\eta^\prime$-meson in the decay
         $\Upsilon (1S) \to \eta^\prime ggg \to \eta^\prime X$ 
         for the three assumed phenomenological models for the 
         $\eta^\prime g^* g$ vertex given in the text, in comparison 
         with the data from the CLEO collaboration~\cite{Artuso:2002px}.
         Note that only the data points for $z=2 
         E_{\eta^\prime}/M \geq 0.7$ are relevant for 
         comparison with the theoretical models.} %
\end{figure}

We start our numerical calculations by reproducing the already known 
results for the $\eta^\prime$-meson energy spectrum 
$dn/dz$~\cite{Artuso:2002px,Kagan:2002dq}, as this provides a good 
consistency check of our calculations. 
For this purpose, the spectrum in Fig.~\ref{fig:CLEO1} 
is calculated with the same set of phenomenological parameters 
as has been used in Ref.~\cite{Kagan:2002dq} for the following 
three input forms\footnote{
We thank Alex Kagan for providing us the input parameters 
used in their analysis of Fig.~\ref{fig:CLEO1}. It should be noted 
that in the rest of our paper we have used the input parameters 
displayed in Table~\ref{tab:input}.}:  
\begin{itemize}

\item[{\bf a)}] A slowly falling EVF:
$F_{\eta^\prime g} (p^2) \simeq 2.1 \, {\rm GeV}^{-1} \, 
[ \alpha_s (p^2) / \alpha_s (m_{\eta^\prime}^2)]$ where 
the two-loop expression is used for the strong coupling~$\alpha_s$.

\item[{\bf b)}] A rapidly falling EVF of the form~(\ref{eq:VF}) 
with the function $H (p^2)$ approximated by the constant value 
$H (p^2) \simeq 1.7$~GeV$^{-1}$.

\item[{\bf c)}] An intermediate EVF: $F_{\eta^\prime g} (p^2) 
\simeq 12.5 \, {\rm GeV}/(p^2 + M_0^2)$ with $M_0 = 2.25$~GeV. 

\end{itemize}
The shapes of the $\eta^\prime$-meson energy spectrum resulting 
from these EVFs are presented in Fig.~\ref{fig:CLEO1}.
%
%
%
%
The experimental points in the decay $\Upsilon(1S) \to \eta^\prime X$ 
shown in this figure are taken from Ref.~\cite{Artuso:2002px}.
Our results are in agreement with the ones in Ref.~\cite{Kagan:2002dq}, 
and we confirm the previous observation that the curve~(b) corresponding 
to the rapidly falling EVF is in reasonable agreement with 
the experimental data in the hard-energy region $z \ge 0.7$
\cite{Artuso:2002px}. Note that the allowed kinematic domain of the 
variable~$z$ is $z_0 < z < 1 + z_0^2/4$, where $z_0 = 2 m_{\eta^\prime}/M$. 
The region $1 < z < 1 + z_0^2/4$  is too small to be resolved
experimentally, given the statistics of the CLEO data~\cite{Artuso:2002px}, 
and hence the energy distribution above~$z=1$ is included in the last
energy bin $0.9 < z < 1.0$.

\begin{figure}[tb]
\centerline{
\psfig{file=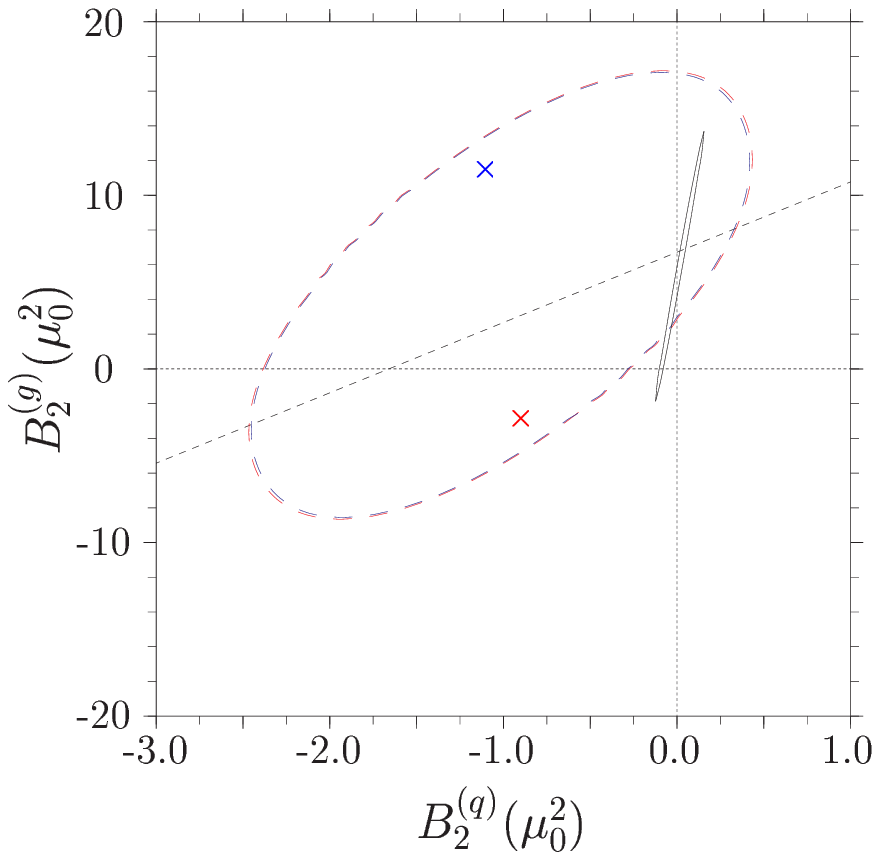,width=.45\textwidth} 
\qquad 
\psfig{file=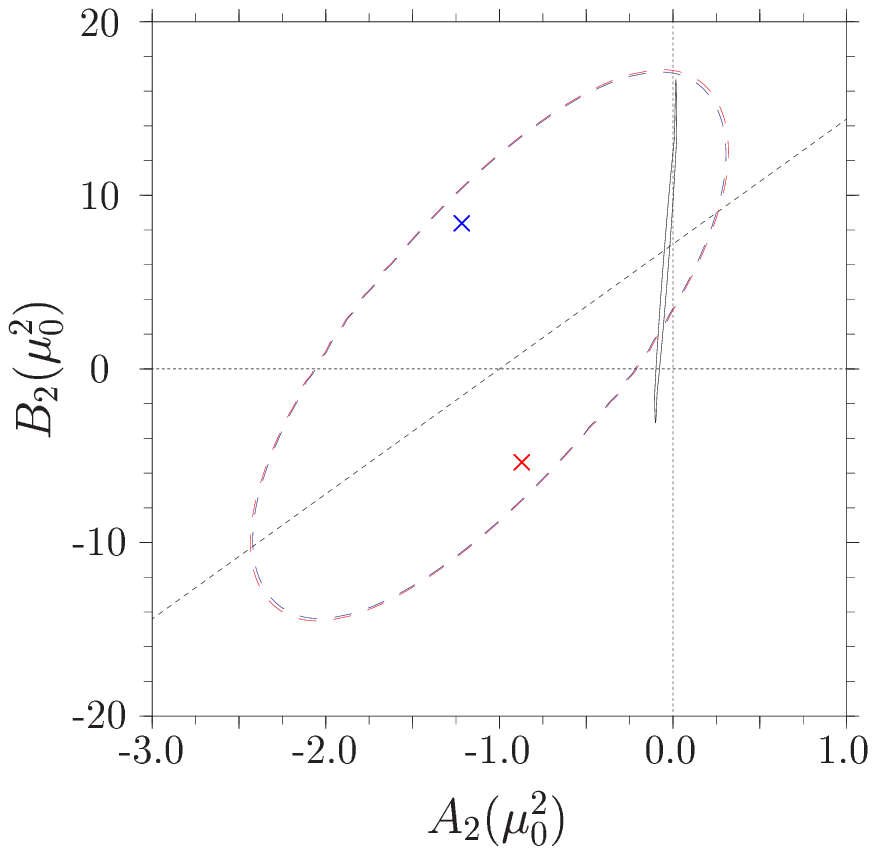,width=.45\textwidth} 
}
\caption{\label{fig:chi2}%
         The $\pm 1~\sigma$ contours (long-dashed curves) in the parameter 
         space $B_2^{(g)}(\mu_0^2) - B_2^{(q)}(\mu_0^2)$ (left frame) 
         and for the Gegenbauer moments $B_2(\mu_0^2) - A_2(\mu_0^2)$ 
         (right frame) resulting from the analysis of the last three 
         experimental bins in the $\eta^\prime$-meson energy spectrum 
         in the process $\Upsilon(1S) \to \eta^\prime X$ measured by 
         the CLEO collaboration~\cite{Artuso:2002px}. The crosses 
         ($\times$) represent the two solutions with the minimum $\chi^2$
         given in the text.
         The narrow solid contours result from the 
         analysis of the data on the $\eta^\prime - \gamma$ transition 
         form factor, scaled from Ref.~\cite{Kroll:2002nt}.  
         The short-dashed lines in both the frames result by demanding 
         that the vertex function $F_{\eta^\prime g}(p^2)$ remains 
         positive for all values of $p^2 > m_{\eta^\prime}^2$, allowing 
         only the regions below these lines.  Note that all 
         the contours correspond to using $\mu_0^2 = 2~{\rm GeV}^2$.
} 
\end{figure}

Let us now turn to the analysis of the perturbative-QCD based  
form of the effective vertex 
presented in Eqs.~(\ref{eq:VF})--(\ref{eq:G2g}). To fit the 
parameters~$B_2^{(q)}(\mu_0^2)$ and~$B_2^{(g)}(\mu_0^2)$ from the existing 
experimental 
data on the $\Upsilon (1S) \to \eta^\prime g g g \to \eta^\prime X$ decay, 
it is convenient to first derive an approximate numerical  
formula for the  $\eta^\prime$-meson energy spectrum. This  
expression  will come in handy for subsequent comparison with                      
other independent calculations and will also facilitate undertaking  
analysis of the data in future by the experimental groups themselves.
%
 As the amplitude of the process 
considered is linear in~$B_2^{(q)}$ and~$B_2^{(g)}$, the energy 
spectrum $dn/dz$ is quadratic in these parameters. The general form for 
the energy spectrum in the decay $\Upsilon (1S) \to 
\eta^\prime g g g \to \eta^\prime X$ can be written as follows
(suppressing in this equation the scale-dependence of the coefficients 
for ease of writing):
\begin{eqnarray}
\frac{d \tilde n}{dz} (z, B_2^{(q)}, B_2^{(g)}) & = & 
C_{00} (z) + C_{0q} (z) B_2^{(q)} + C_{0g} (z) B_2^{(g)} 
\label{eq:MC-interpolation} \\ 
& + & C_{qq} (z) [B_2^{(q)}]^2 + 
C_{qg} (z) B_2^{(q)} B_2^{(g)} + C_{gg} (z) [B_2^{(g)}]^2 .
\nonumber
\label{eq:dndz} 
\end{eqnarray}
We have generated the theoretical $E_{\eta^\prime}$-spectrum, using a 
high statistics Monte Carlo phase space programme, with fixed values 
of the coefficients~$B_2^{(q)}(\mu_0^2)$ and~$B_2^{(g)}(\mu_0^2)$, and 
have varied their values over a certain range  
to numerically determine the dependence  of the spectrum on these 
coefficients. Other parameters in our numerical analysis are listed in 
Table~\ref{tab:input}, which are the same as the ones used in the 
analysis of the $\eta^\prime - \gamma$ and~$\eta - \gamma$ transition
form factors~\cite{Kroll:2002nt}, for the sake of consistency, 
as we are going  to make use of this analysis.  
\begin{table}[tb]
\caption{Input parameters used in the numerical analysis.}
\label{tab:input}
\begin{center}
\begin{tabular}{ll}
$M = $               9.46~GeV  & $m_c = $ 1.3~GeV \\
$m_{\eta^\prime} = $ 958~MeV   & $m_b = $ 4.3~GeV \\
$f_\pi = $           133~MeV   &
$\Lambda_{\rm QCD}^{(4)} = $ 305~MeV \\
$\mu_0^2 = $         2~GeV$^2$ &
\end{tabular}
\end{center}
\end{table}
We have not included any errors on $m_c$, $m_b$ and 
$\Lambda_{\rm QCD}^{(4)}$, as the parametric uncertainties 
from the Gegenbauer coefficients are by far the largest, 
which we study.
\begin{table}[tb]
\caption{The coefficients in the interpolating function $d\tilde n/dz$
         defined in the text for the decay $\Upsilon(1S) \to 
\eta^\prime X$. The numbers in the brackets
         are from the Monte Carlo statistical errors.}
\label{tab:coeff}
\medskip
\hspace*{-2mm}
\begin{tabular}{|l|l|l|l|l|l|l|}
\hline
$z$ & $C_{00} (z)$ & $C_{0q} (z)$ & $C_{0g} (z)$ & $C_{qq} (z)$ &
$C_{qg} (z)$ & $C_{gg} (z)$
\\ \hline
$0.6 \div 0.7$ &
1.9172(27) & 2.5187(36) & -0.3108(13) & 0.8744(19) &
-0.1888(07) & 0.0146(06)
\\
$0.7 \div 0.8$ &
1.6750(25) & 2.2368(34) & -0.2583(12) & 0.7953(18) &
-0.1560(07) & 0.0122(06)
\\
$0.8 \div 0.9$ &
1.2768(21) & 1.7344(29) & -0.1855(10) & 0.6343(15) &
-0.1097(06) & 0.0089(05)
\\
$0.9 \div 1.0$ &
0.6475(17) & 0.8945(17) & -0.0866(07) & 0.3396(09) &
-0.0484(05) & 0.0044(05)
\\
\hline
\end{tabular}
\end{table}
The coefficients $C_{ab} (z)$ [$a, b = 0, q, g $] for the 
four bins of the $\eta^\prime$-meson energy spectrum are presented in
Table~\ref{tab:coeff}. The numbers presented in the parentheses represent
the statistical error of our Monte Carlo calculations, for which we have 
used the Monte Carlo phase space generator FOWL from the CERN Library of 
programmes. 
With the help of the program MINUIT~\cite{James:1975dr},
the following best fit values of the parameters~$B_2^{(q)}(\mu_0^2)$ 
and~$B_2^{(g)}(\mu_0^2)$ are obtained: 
\begin{eqnarray}
B_2^{(q)} (\mu_0^2) = -0.89^{+ 1.32}_{- 1.58}, &&
B_2^{(g)} (\mu_0^2) = -2.86^{+ 20.04}_{- 5.80}, \qquad
\chi^2 = 2.45 ,
\label{eq:fit-3bin} \\
B_2^{(q)} (\mu_0^2) = -1.09^{+ 1.51}_{- 1.36}, &&
B_2^{(g)} (\mu_0^2) = 11.53^{+ 5.55}_{- 20.09}, \qquad
\chi^2 = 2.37 ,
\nonumber
\end{eqnarray}
for the last three experimental bins with $z > 0.7$ (the 
stated $\chi^2$ corresponds to three degrees of freedom), and
\begin{eqnarray}
B_2^{(q)} (\mu_0^2) = -0.77^{+ 0.73}_{- 0.78}, &&
B_2^{(g)} (\mu_0^2) = -4.36^{+ 6.28}_{- 4.46}, \qquad
\chi^2 = 24.13 ,
\label{eq:fit-4bin} \\
B_2^{(q)} (\mu_0^2) = -1.29^{+ 0.76}_{- 0.73}, &&
B_2^{(g)} (\mu_0^2) = 12.51^{+ 4.53}_{- 6.33}, \qquad
\chi^2 = 23.69 ,
\nonumber
\end{eqnarray}
for the four experimental data points in the large-$z$ region ($z>0.6$).
The minimum~$\chi^2$ of the fit in the last case, namely 
$\chi^2 \simeq 24$ for four degrees of freedom, is unacceptably large. 
Thus, as already observed 
in Ref.~\cite{Artuso:2002px}, only the last three bins in the energy 
spectrum are dominated by the $\Upsilon (1S) \to \eta^\prime g g g \to 
\eta^\prime X$ decay. Following this, we concentrate on the analysis 
of the last three bins with $z \ge 0.7$.   
The $1\sigma$ contours both in the Gegenbauer 
coefficients ($B_2^{(q)}(\mu_0^2)$, $B_2^{(g)}(\mu_0^2)$) and in the 
Gegenbauer moments ($A_2 (\mu_0^2)$, $B_2 (\mu_0^2)$) are presented 
as long-dashed curves in Fig.~\ref{fig:chi2}.  
%
%
%
%
%
The best fit values~(\ref{eq:fit-3bin}) are denoted by the 
crosses~($\times$). 
The narrow contours (solid curves) also shown in these figures result 
from the analysis of the data on the $\eta^\prime - \gamma$ transition 
form factor~\cite{Kroll:2002nt}. As we are using 
$\mu_0^2 = 2$~GeV$^2$ in our analysis, the results from 
Ref.~\cite{Kroll:2002nt} have been rescaled from $\mu_0^2 = 1$~GeV$^2$ 
to $\mu_0^2 = 2$~GeV$^2$ 
with the help of Eqs.~(\ref{eq:An}) and~(\ref{eq:Bn}) for $n = 2$.
We note that the analysis of the $\eta^\prime - \gamma$ data, being more
sensitive to the quark-antiquark LCDA of the $\eta^\prime$-meson, provides 
a much more precise determination of the parameter $B_2^{(q)} (\mu_0^2)$  
and hence of the Gegenbauer moment~$A_2 (\mu_0^2)$, than what can be
determined at present from the CLEO data on $\Upsilon(1S) \to 
\eta^\prime X$. However, what concerns the
coefficient $B_2^{(g)} (\mu_0^2)$, and also the Gegenbauer moment 
$B_2 (\mu_0^2)$, the CLEO data on $\Upsilon(1S) \to \eta^\prime X$, 
despite its statistical limitations, has cut out some of the 
allowed parameter region from the $\eta^\prime - \gamma$ analysis. 
The additional constraint that follows from demanding that the vertex 
function $F_{\eta^\prime g} (p^2)$ remains positive definite for all 
values of $p^2 > m_{\eta^\prime}^2$ is shown through the short-dashed 
lines in these figures, admitting only the parameter space below these 
lines. We will discuss this constraint in more detail below.

\begin{figure}[tb]
\centerline{
\psfig{file=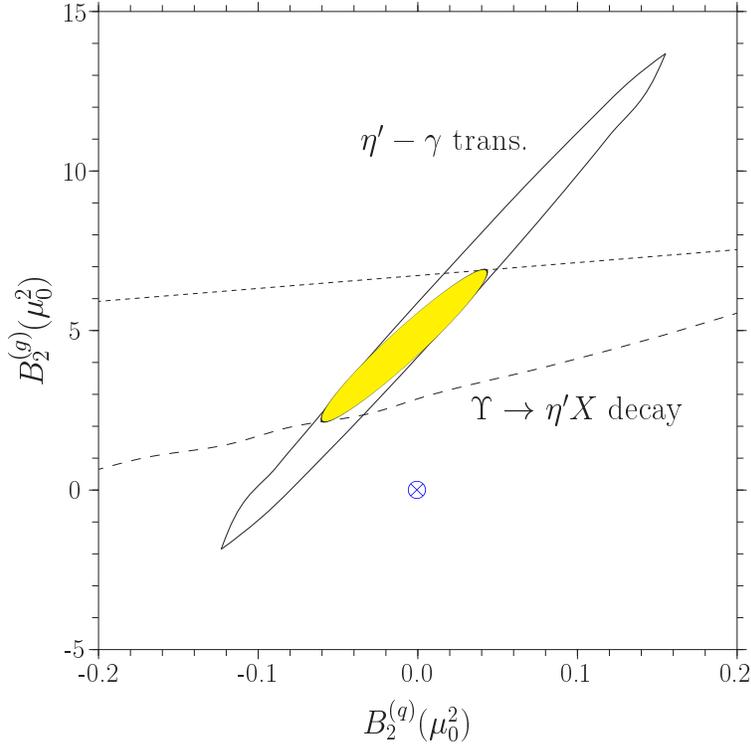,width=.65\textwidth} 
}
\caption{\label{fig:chi2B2-bf}%
   The combined fit for the parameters~$B^{(q)}_2 (\mu_0^2=2~{\rm GeV}^2)$ 
   and~$B^{(g)}_2 (\mu_0^2=2~{\rm GeV}^2)$ from the data on the 
   $\eta^\prime - \gamma$ transition form factor (solid curve) and 
   $\Upsilon (1S) \to \eta^\prime X$ decay (long-dashed and 
   short-dashed curves) with the 
   requirement that the vertex $F_{\eta^\prime g} (p^2)$ 
   remains positive definite in the entire~$p^2$ region. 
   The resulting 1$\sigma$ contour (combined best fit) is shown by 
   the yellow (shaded) region. The point denoted by $\bigotimes$ 
   corresponds 
   to the asymptotic light-cone distribution amplitude.}  
\end{figure}

A blow up of the overlapping region in the parameter space 
($B_2^{(g)} (\mu_0^2), B_2^{(q)} (\mu_0^2)$)
is shown in Fig.~\ref{fig:chi2B2-bf}. 
%
%
%
%
%
Here, the $1\sigma$ contours following from the $\eta^\prime - \gamma$ 
transition form factor analysis and the $\Upsilon (1S) \to \eta^\prime 
X$ decay fit are shown through the solid and dashed curves, 
respectively. In drawing the allowed parameter space from the 
$\Upsilon (1S) \to \eta^\prime  X$ decay fit, we have imposed the
additional condition that the $\eta^\prime g^* g$ vertex function  
$F_{\eta^\prime g} (p^2)$ for $p^2 > m_{\eta^\prime}^2$ has the same 
sign as the corresponding function calculated with 
the asymptotic forms of the $\eta^\prime$-meson quark-antiquark and 
gluonic LCDAs which is defined as positive-definite in the entire~$p^2$
region. With Eq.~(\ref{eq:VF}) taken into account, this condition implies 
$H (p^2) > 0$ and results [with the help of the explicit 
from~(\ref{eq:H-HSA}) of the function~$H (p^2)$] in the following 
inequalities:
\begin{eqnarray} 
B_2 (\mu_0^2) & \le & \frac{36}{5} \left [ 1 + A_2 (\mu_0^2) \right ] ,  
\label{eq:condition} \\ 
B_2^{(g)} (\mu_0^2) & \le & 
\frac{36 + (36 - 5 \rho_2^{(q)}) \, B_2^{(q)} (\mu_0^2)}
{5 - 36 \rho_2^{(g)}} ,  
\nonumber 
\end{eqnarray} 
where the function~$G_2^{(g)} (1, \zeta)$~(\ref{eq:G2g}) is 
approximated by its value~$1/6$ in the large-$|p^2|$ asymptotics.  
The values $\rho_2^{(q)} = 2.86$ and $\rho_2^{(g)} = - 0.01$ were 
taken for the constrained parameters in the numerical analysis.   
The positivity constraint  removes the larger values of $B_2^{(g)} 
(\mu_0^2)$ above the short-dashed curve in Fig.~\ref{fig:chi2B2-bf}, 
which would otherwise force the vertex function to cross the zero
at some value of $p^2$ and become negative. This is exemplified below
for specific values of the Gegenbauer coefficients.
 The resulting combined best fit of the 
parameters~$B^{(q)}_2 (\mu_0^2)$ and~$B^{(g)}_2 (\mu_0^2)$ is shown 
as the coloured (shaded) region. This yields the following correlated 
values: 
\begin{eqnarray}
B_2^{(q)} (\mu_0^2) = -0.008 \pm 0.054, \qquad 
B_2^{(g)} (\mu_0^2) = 4.6 \pm 2.5,   
\label{eq:combine-fit} \\  
A_2 (\mu_0^2) = - 0.054 \pm 0.029, \qquad\hspace*{2mm} 
B_2 (\mu_0^2) = 4.6 \pm 2.7, 
\hspace*{3mm} 
\nonumber 
\end{eqnarray}
with the central values having $\chi^2 = 2.66$ for three degrees of 
freedom. For comparison, the point shown as~$\bigotimes$~, 
corresponding to the asymptotic LCDA, has $\chi^2 = 8.41$ for three 
degrees of freedom. Thus, the asymptotic form 
of the $\eta^\prime$-meson LCDA is not favoured by our analysis.
%
%
%
\begin{figure}[tb]
\centerline{
\psfig{file=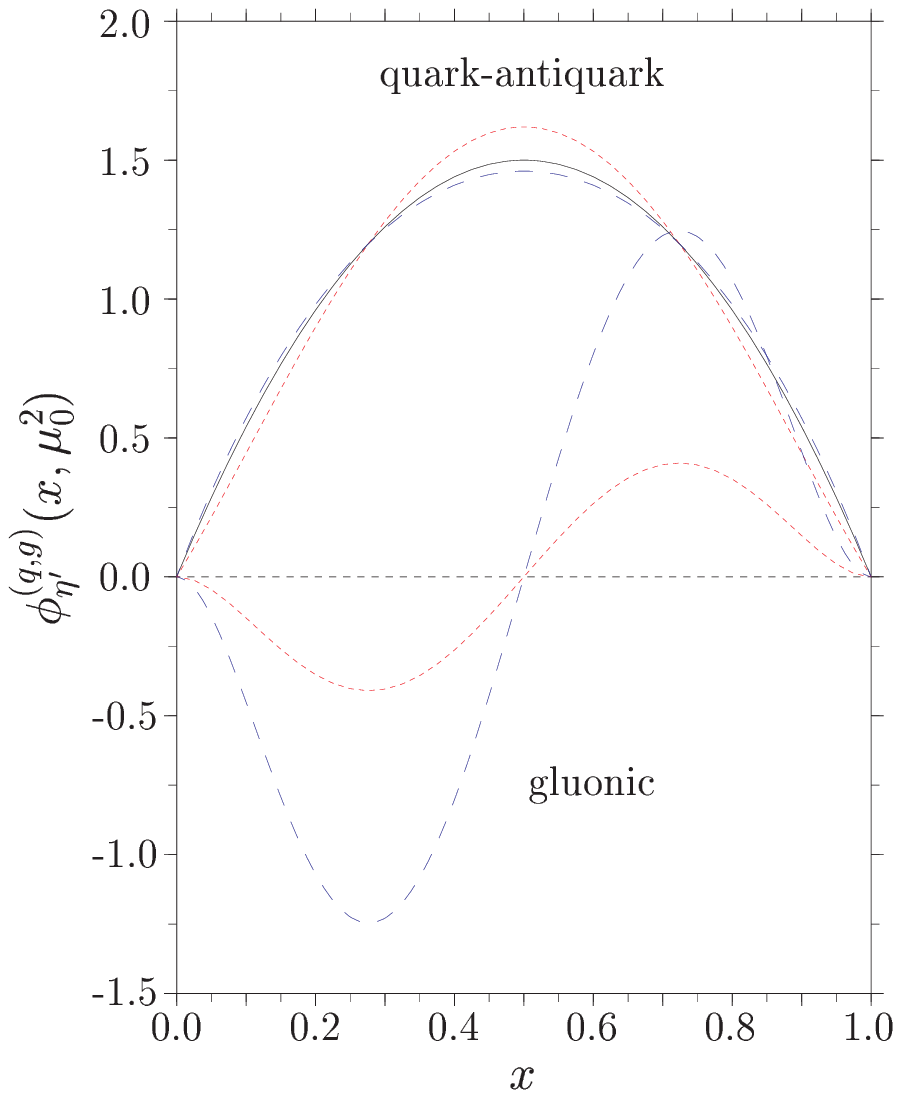,width=.45\textwidth}
\quad
\psfig{file=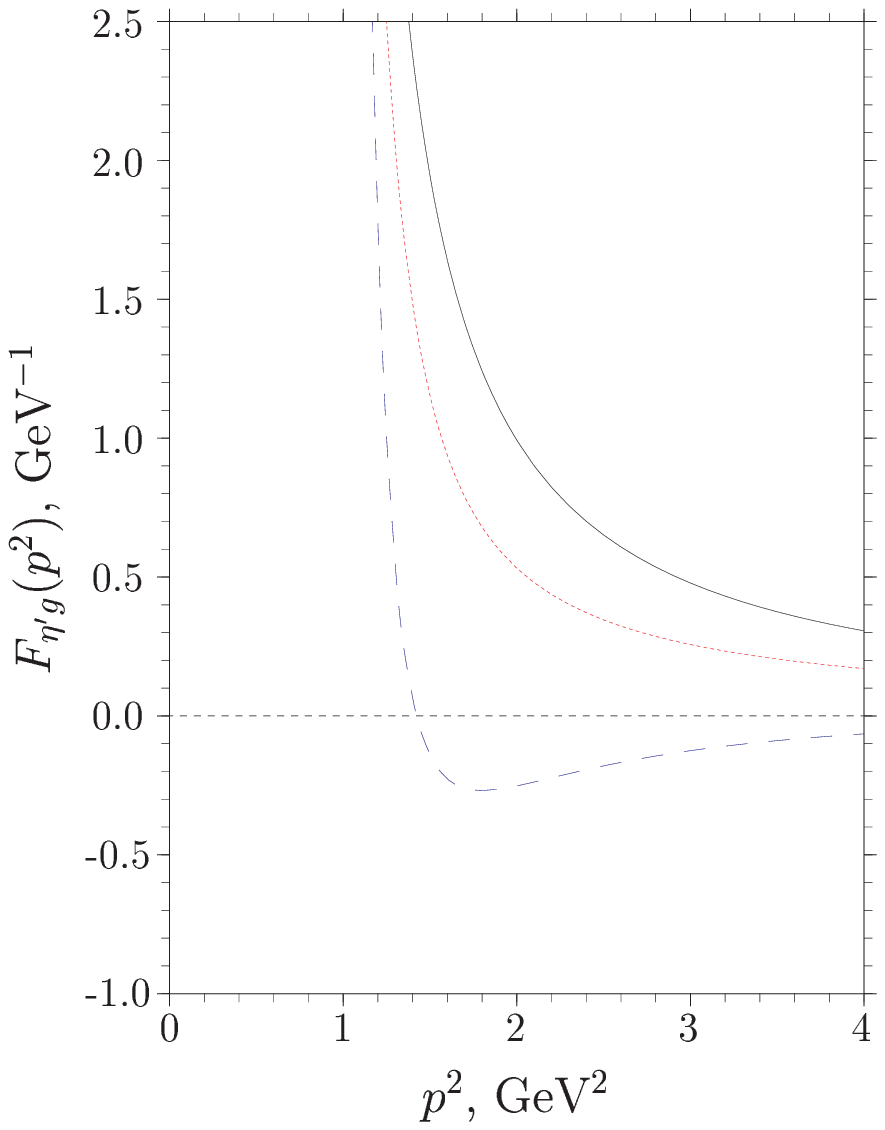,width=.45\textwidth}
}
\caption{\label{fig:LCDAs}%
      The $\eta^\prime$-meson quark-antiquark and 
      gluonic LCDAs $\phi^{(q,g)}_{\eta^\prime} (x,\mu_0^2)$ as a 
      function of~$x$ (left frame), and the resulting $\eta^\prime - g$ 
      transition form factor $F_{\eta^\prime g} (p^2)$ (right frame). 
      The solid curves correspond to the asymptotic 
      quark-antiquark LCDA, while the LCDAs for the combined best fit 
      values of the Gegenbauer coefficients given 
      in Eq.~(\ref{eq:combine-fit}) are drawn as 
      dotted curves. The LCDAs with the values $B^{(q)}_2 = 0.15$ 
      and $B^{(g)}_2 = 13.5$, which are allowed within~$1\sigma$ 
      from the analysis of the data on the 
      $\eta^\prime - \gamma$ transition form factor~\cite{Kroll:2002nt}, 
      are presented as the dashed curves. Note, that for this case,
      the function $F_{\eta^\prime g} (p^2)$ is no longer 
      positive-definite, as shown in the right frame.} 
\end{figure}
%
%

The shapes of the quark-antiquark and gluonic LCDAs are presented in
Fig.~\ref{fig:LCDAs} (left frame); the resulting $\eta^\prime - g$ 
transition form factor $F_{\eta^\prime g} (p^2)$ corresponding 
to these LCDAs is also shown in this figure (right frame).
The solid and dotted curves in this figure correspond to the 
asymptotic LCDA and the combined best fit values of the Gegenbauer 
coefficients~(\ref{eq:combine-fit}), respectively. 
We also show, for the sake of illustration,   
representative LCDAs with the Gegenbauer coefficients having the values
$B^{(q)}_2 (2~{\rm GeV}^2)= 0.15$ and $B^{(g)}_2 (2~{\rm GeV}^2) = 13.5$. 
These parametric values are taken from the analysis in 
Ref.~\cite{Kroll:2002nt}, but are in conflict  with the positive 
definiteness of the vertex function $F_{\eta^\prime g} (p^2)$, 
as shown by the dashed curve for $F_{\eta^\prime g} (p^2)$. 
In fact, the constraint of positivity on the effective vertex function 
cuts out regions in the parameter space $B^{(q)}_2 (2~{\rm GeV}^2) > 0.045$ 
and $B^{(g)}_2 (2~{\rm GeV}^2) > 7.1$, as otherwise the contribution 
from the gluonic LCDA starts to dominate, which makes the
vertex function $F_{\eta^\prime g} (p^2)$ negative for some range of~$p^2$ 
considered here and in Ref.~\cite{Kroll:2002nt}. Thus, positivity 
criterion provides an effective constraint on the magnitude
of the coefficient $B^{(g)}_2 (\mu_0^2)$, reducing significantly the 
otherwise allowed range in Eq.~(\ref{eq:B2-fit}). In the context of our 
analysis, we note that the asymptotic $\eta^\prime$-meson LCDA provides  
a fair description (though not the best fit) of the current data on 
$\Upsilon(1S) \to ggg^* (g^* \to \eta^\prime g) \to \eta^\prime X$, 
and hence one anticipates that the subasymptotic corrections in the
LCDAs, and the vertex function $F_{\eta^\prime g} (p^2)$, while important 
in the analysis of data, are not dominant.\footnote{The approximate 
validity of the asymptotic transition form factors involving the 
$\pi$-, $\eta$- and $\eta^\prime$-mesons compared to the data, 
observed in Refs.~\cite{Feldmann:1999uf,Feldmann:2002kz}, also 
suggests that the coefficient $B_2^{(g)} (\mu_0^2)$ in the
$\eta^\prime$-meson is bounded by this data. The gluonic components  
of the $\pi$- and the $\eta$-meson LCDAs are small in any case.}

\begin{figure}[tb]
\centerline{
\psfig{file=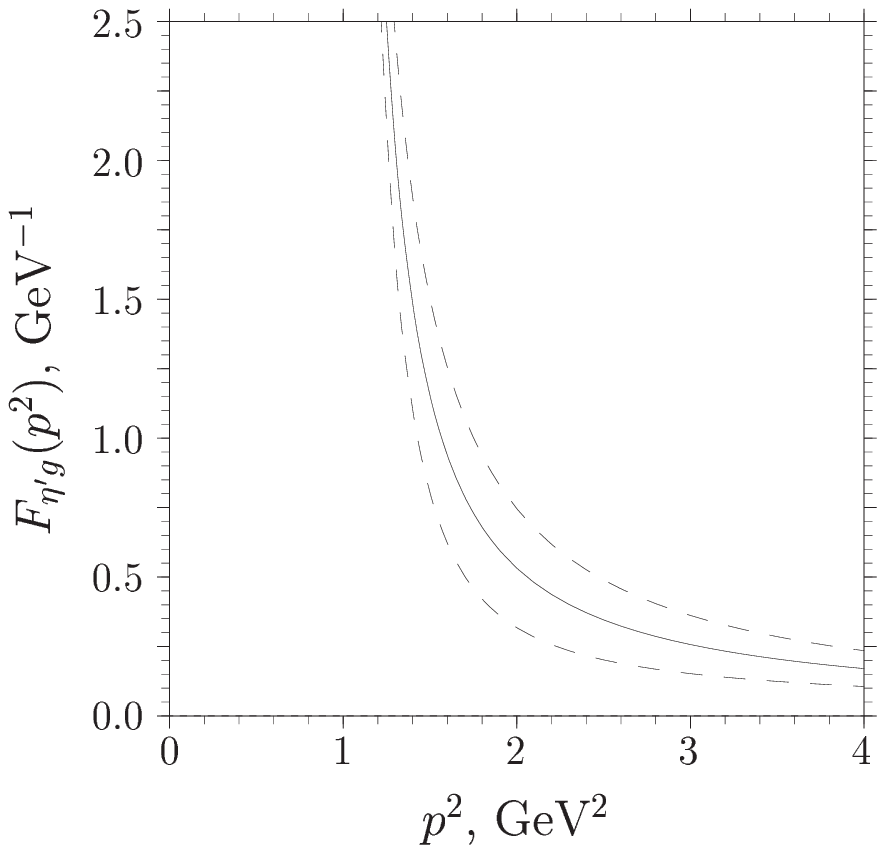,width=.415\textwidth} 
\qquad 
\psfig{file=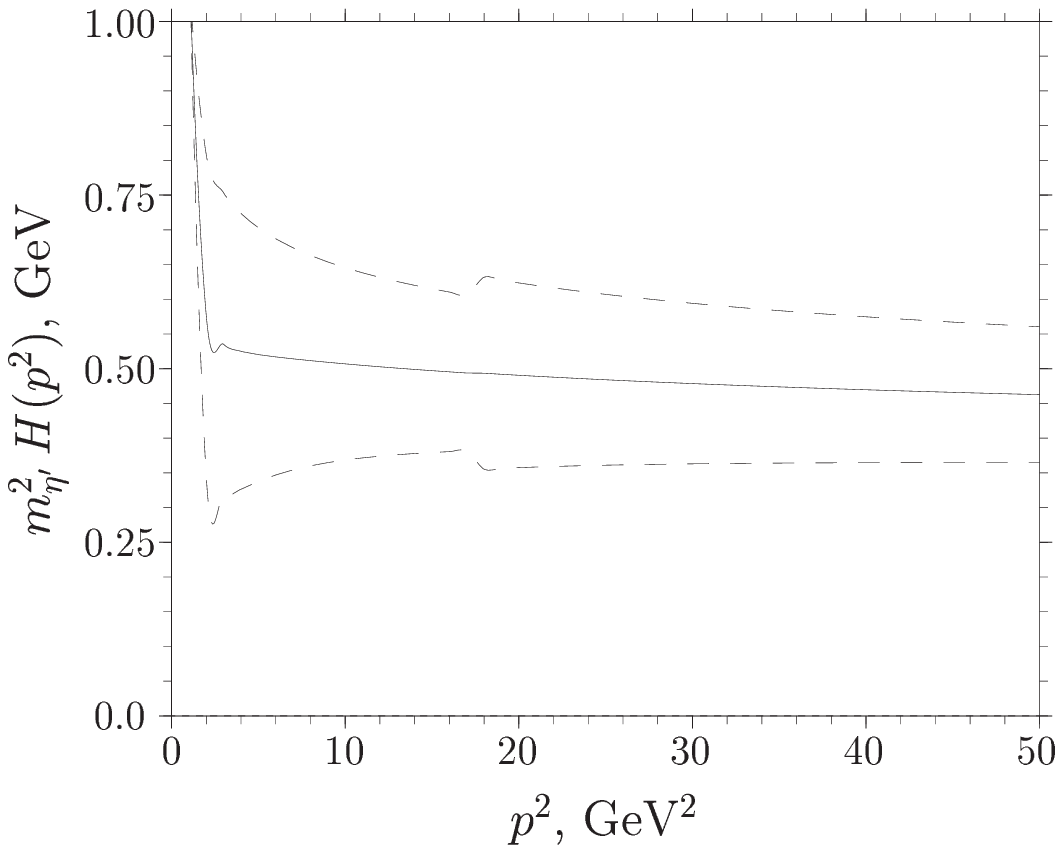,width=.485\textwidth} 
}
\caption{\label{fig:EVF}%
      The $\eta^\prime - g$ transition form factor $F_{\eta^\prime g} 
      (p^2)$, defined in Eq.~(\ref{eq:VF}) (left frame), and the function 
      $m_{\eta^\prime}^2 H (p^2)$, defined in Eq.~(\ref{eq:H-HSA})  
      (right frame), as functions of the gluon virtuality~$p^2$. 
      The various curves correspond to the input values of 
      the Gegenbauer coefficients~$B^{(q)}_2 (\mu_0^2)$
      and~$B^{(g)}_2 (\mu_0^2)$ resulting from the combined best fit 
      shown in Fig.~\ref{fig:chi2B2-bf}. 
      The solid curves correspond to the central values of the fit, 
      while the dashed curves are drawn for the parameters 
      corresponding to the uppermost and lowermost allowed values of 
      the combined best fit contour shown in Fig.~\ref{fig:chi2B2-bf}.} 
\end{figure}

\begin{figure}[tbh]
\centerline{
\psfig{file=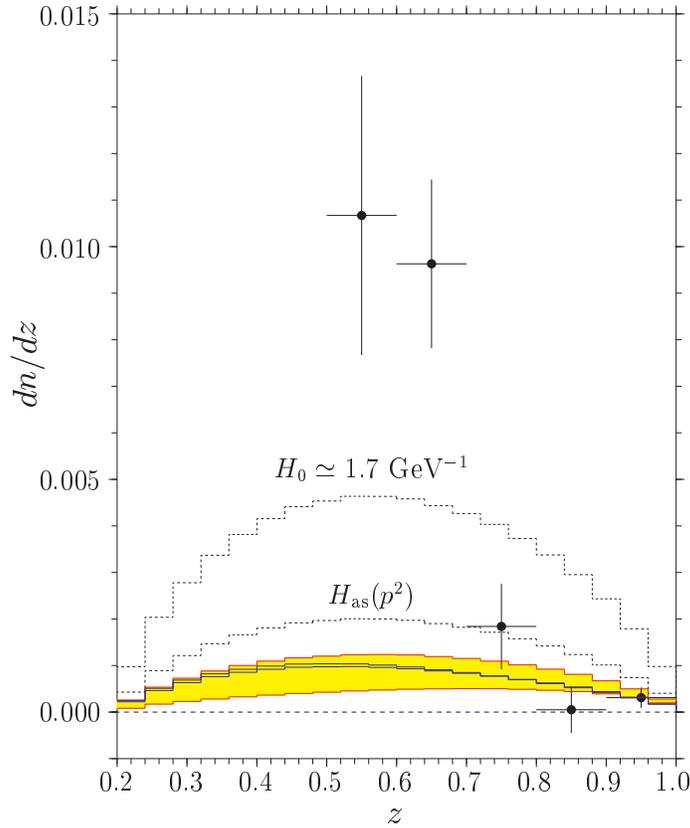,width=.60\textwidth}
}
\caption{\label{fig:CLEO2}%
         Energy spectrum of the $\eta^\prime$-meson in the
         decay $\Upsilon (1S) \to \eta^\prime X$. 
         The upper dotted curve corresponds to a constant
         value of the function $H (p^2) = H_0 \simeq 1.7$~GeV$^{-1}$, 
         and the curve marked as $H_{\rm as} (p^2)$ corresponds to the 
         asymptotic form of the
         $\eta^\prime$-meson LCDA (i.e., $B^{(q)}_2 = 0$ and
         $B^{(g)}_2 = 0$). The spectrum with the Gegenbauer coefficients
         in the combined best-fit range of these parameters is shown 
         by the yellow (shaded) region. The blue (solid) curve inside 
         this region corresponds to the best fit values of the 
         Gegenbauer coefficients from the analysis of the $\Upsilon (1S) 
         \to  \eta^\prime X$ data alone.}  
\end{figure}

The EVF $F_{\eta^\prime g} (p^2)$ and the function 
$m_{\eta^\prime}^2 H (p^2)$ for the combined best fit 
values~(\ref{eq:combine-fit}) are presented in Fig.~\ref{fig:EVF}. 
%
%
%
%
%
The $\eta^\prime$-meson energy spectrum for the EVF motivated by the
perturbative QCD analysis is presented in Fig.~\ref{fig:CLEO2} in
comparison with the experimental data~\cite{Artuso:2002px} and the
spectrum corresponding to the ``rapidly falling'' form of the
EVF~\cite{Kagan:2002dq} labeled by the constant value of the 
function $H (p^2) = H_0 \simeq 1.7$~GeV$^{-1}$. The second
dotted line labeled by $H_{\rm as} (p^2)$ corresponds to the EVF
when only the asymptotic form of the $\eta^\prime$-meson LCDA
($B^{(q)}_2 = 0$ and $B^{(g)}_2 = 0$) is taken into account.
The yellow (shaded) region demarcates the spectrum with the Gegenbauer 
coefficients having values in the range of the combined best 
fit~(\ref{eq:combine-fit}). The blue (solid) curve lying inside this 
region 
corresponds to the best fit values~(\ref{eq:fit-3bin}) obtained from 
the analysis of the $\Upsilon (1S) \to \eta^\prime X$ decay only. 
From Fig.~\ref{fig:CLEO2} it is seen that the rapidly falling 
phenomenological EVF with $H_0 = 1.7~{\rm GeV}^{-1}$ gives 
a harder $\eta^\prime$-meson energy spectrum for the 
large energy~$z$ region compared to the CLEO data.
The spectrum with the asymptotic form of the $\eta^\prime$-meson 
LCDA is well correlated with the experimental point in the bin 
$0.7 \le z \le 0.8$, but overestimates the data in the other two bins.  
It should be noted that the last two bins (especially 
the bin $0.8 \le z \le 0.9$) are very uncertain in the current data,
which has to be statistically improved to draw sharper conclusions.
%
%
%
%

\section{Summary} 
\label{sec:conclusion} 

We have calculated the $\eta^\prime$-meson energy spectrum in the 
decay $\Upsilon (1S) \to \eta^\prime g g g \to \eta^\prime X$ in 
leading order perturbative QCD in the static quark limit for the 
Orthoquarkonium.
Assuming some phenomenological vertex functions, our results are in 
agreement with the ones obtained earlier in Ref.~\cite{Kagan:2002dq}. 
In our analysis, the $\eta^\prime$-meson is described in the  
leading-twist (twist-two) accuracy in terms of the quark-antiquark 
and gluonic LCDAs for which the asymptotic and the first 
non-asymptotic terms are taken into account. In this approximation,   
the $\eta^\prime g^* g$ EVF depends essentially on the 
Gegenbauer coefficients~$B^{(q)}_2 (\mu_0^2)$ and~$B^{(g)}_2 (\mu_0^2)$.
They are determined from the CLEO data on $\Upsilon (1S) \to \eta^\prime 
X$ in the large-$z$ region ($z \ge 0.7$) of the $\eta^\prime$-meson 
energy spectrum~\cite{Artuso:2002px}, which is well explained by our 
perturbative QCD approach. However, the resulting  $1\sigma$ contour in 
the Gegenbauer coefficients have a large dispersion. Combining the 
analysis of the $\Upsilon (1S) \to \eta^\prime X$ data with an earlier 
analysis of the $(\eta^\prime, \eta) - \gamma$ transition from factors 
in a similar theoretical framework~\cite{Kroll:2002nt}, and requiring 
additionally that the vertex function $F_{\eta^\prime g} (p^2)$ 
remains positive definite in the entire~$p^2 > m_{\eta^\prime}^2$ region, 
yield much improved determination of the Gegenbauer coefficients, yielding  
$B^{(q)}_2 (2~{\rm GeV}^2) = -0.008 \pm 0.054$ and 
$B^{(g)}_2 (2~{\rm GeV}^2)= 4.6 \pm 2.5$. Our analysis improves the
phenomenological profile of the LCDAs in the $\eta^\prime$-meson, 
and in turn yields a better determination of the vertex function
$F_{\eta^\prime g} (p^2)$ compared to the earlier estimates of 
the same. The resulting function (the $\eta^\prime$ -- gluon transition 
form factor) is presented in this paper including the $\eta^\prime$-meson 
mass effects.

\section{Acknowledgments}
\label{sec:acknowledgments}

We would like to thank Alex Kagan, Christoph Greub, Peter Minkowski and
Sheldon Stone for numerous helpful discussions. We also thank Gustav 
Kramer for his comments on the earlier version of this manuscript. The 
work of A.Ya.P. has been supported by the Schweizerischer Nationalfonds. 

\appendix
\section{Matrix Element Squared for the Decay 
        $\Upsilon(1S) \to ggg^* (g^* \to \eta^\prime g)$}
\label{sec:A1}

In Sec.~\ref{sec:Branch}, we have displayed explicitly the terms 
${\cal M}_i$ ($i = 1, 2, 3$) in the decomposition of the total 
amplitude of the decay $\Upsilon (1S) \to \eta^\prime g g g$ in 
the leading-order perturbative QCD [Eqs.~(\ref{eq:ME1})--(\ref{eq:ME3})]. 
Squaring this amplitude results in very lengthy expressions. 
Nevertheless, it is possible to present compact 
analytical expressions for some parts of the matrix element squared. 
In particular, the square of the term~${\cal M}_1$, summed over the 
polarizations and colours of the final gluons and averaged over the 
polarization states of the $\Upsilon$-meson, can be written as follows: 
\begin{equation}
\frac{1}{3} \sum |{\cal M}_1|^2 = \frac{512}{3} \, C_F B_F \pi^3
\alpha_s^3 (M^2) \, \frac{|\psi (0)|^2}{M}
\left | F_{\eta^\prime g} (p_1^2) \right |^2
\frac{J_{11} ({\cal P}, k_i, p_i)} 
     {({\cal P} k_2)^2 ({\cal P} k_3)^2 (p_1^\prime p_1)^2 (p_1^2)^2} ,
\label{eq:M1-sq}
\end{equation}
where the well-known property of the symmetrical constants~$d_{ABC}$ 
of the colour $SU(N_c)$-group has been used: 
\begin{equation} 
\sum_{C, D = 1}^{N_c^2 - 1} d_{ACD} \, d_{BCD} = 
\frac{N_c^2 - 4}{N_c} \, \delta_{AB} \equiv 2 B_F \, \delta_{AB} ,    
\label{eq:SU(n)-relation}
\end{equation} 
and the dynamical function $J_{11} ({\cal P}, k_i, p_i)$ has the form: 
\begin{eqnarray} 
J_{11} ({\cal P}, k_i, p_i) & = & 
4 M^2 (k_1 p_1)^2 \left [ (k_2 k_3)^2 ({\cal P} p_1)^2 
+ (k_2 p_1)^2 ({\cal P} k_3)^2 + (k_3 p_1)^2 ({\cal P} k_2)^2 \right ] 
\label{eq:J11} \\ 
& + & 4 M^2 (p_1^2)^2 (k_2 k_3)^2 
\left [ (k_1 k_2)^2 + (k_1 k_3)^2 \right ] 
\nonumber \\ 
& + & 8 M^2 p_1^2 (p_1 k_1)^2 (k_2 k_3) 
\left [ (k_2 k_3)^2 - (k_2 p_1) (k_3 p_1) \right ] 
\nonumber \\ 
& - & 8 M^2 p_1^2 (p_1 k_1) (k_2 k_3)^2 
\left [ (k_1 k_2) (k_2 p_1) + (k_1 k_3) (k_3 p_1) \right ] 
\nonumber \\ 
& + & 2 M^2 p_1^2 (M^2 - p_1^2) 
\left \{ \left [ (k_1 k_2) (k_3 p_1) - (k_1 k_3) (k_2 p_1) \right ]^2  
- (p_1 k_1)^2 (k_2 k_3)^2  \right \}
\nonumber \\ 
& + & 2 p_1^2 (M^2 - p_1^2) 
\left [ M^2 (k_2 k_3) + 2 ({\cal P} k_2) ({\cal P} k_3) \right ] 
\nonumber \\ 
& \times & 
\left [ p_1^2 (k_1 k_2) (k_1 k_3) - (k_1 p_1) (k_1 k_2) (k_3 p_1) 
- (k_1 p_1) (k_1 k_3) (k_2 p_1)  \right ] . 
\nonumber 
\end{eqnarray}
It is easy to check that this function is invariant under 
the interchange $(k_2, p_2) \leftrightarrow (k_3, p_3)$, 
which reflects the Bose symmetry for the set of the Feynman diagrams
considered. The other two dynamical functions~$J_{22} ({\cal P}, k_i, p_i)$ 
and~$J_{33} ({\cal P}, k_i, p_i)$ originating from the squares of the 
terms~${\cal M}_2$ and~${\cal M}_3$, respectively, can be obtained 
from~$J_{11} ({\cal P}, k_i, p_i)$ by the obvious replacements: 
$(k_1, p_1) \leftrightarrow (k_2, p_2)$ and 
$(k_1, p_1) \leftrightarrow (k_3, p_3)$. 

The differential width of the $1 \to 4$ decay has a non-trivial 
dependence on five variables (three angles can be integrated out 
trivially) in the rest frame of the decaying particle. In the Monte 
Carlo generator, the matrix element squared is rewritten in terms of 
the following dimensionless variables: 
\begin{equation} 
x_i = \frac{2 ({\cal P} k_i)}{M^2} , \qquad 
y_i = \frac{p_i^2}{M^2} , \qquad i = 1, 2, 3, 
\label{eq:MC-variables}
\end{equation} 
which satisfy the relation: 
\begin{equation} 
x_1 + x_2 + x_3 + y_1 + y_2 + y_3 = 2 + z_0^2/4 , 
\label{eq:var-relation}
\end{equation} 
where $z_0 = 2 m_{\eta^\prime}/M $ is the relative mass of 
the $\eta^\prime$-meson. 

The experimentally measured quantity is the $\eta^\prime$-meson 
energy spectrum~-- the number of $\Upsilon \to \eta^\prime X$ 
events~$\Delta n_i$ inside the energy bin with the central 
value~$z_i$ and the width~$\Delta z$. This can be calculated 
theoretically in the Monte Carlo approach, which we have used.  
In particular, as an input for the calculation of the contribution 
coming from~$|{\cal M}_1|^2$, the following equation was used: 
\begin{eqnarray}
\frac{\Delta n^{(11)}_i}{\Delta z} & = & 
\hspace*{-4mm} 
\int\limits_{z_i - \Delta z/2}^{z_i + \Delta z/2} \!\! dz \, 
\delta (z - 2 E_{\eta^\prime} / M)  
\int \frac{d{\bf k}_1}{2 \omega_1} \, \frac{d{\bf k}_2}{2 \omega_2} \,
\frac{d{\bf k}_3}{2 \omega_3} \, 
\frac{d{\bf p_{\eta^\prime}}}{2 E_{\eta^\prime}} \,
\delta^{(4)} ({\cal P} - k_1 - k_2 - k_3 - p_{\eta^\prime})  
\nonumber \\
& \times &
\frac{|M F_{\eta^\prime g} (y_1)|^2}{64 \pi^5 (\pi^2 - 9) M^4} \, 
\frac{{\cal F}_{11} (x_i, y_i, \mu)}
     {[y_1 x_2 x_3 (2 - 2 y_1 -  x_2 - x_3)]^2} ,   
\label{eq:dn11} 
\end{eqnarray} 
where the representation~(\ref{eq:M1-sq}) for $|{\cal M}_1|^2$ 
was put into the definition of $dn/dz$~(\ref{eq:BR-spectrum}). 
The dimensionless functions~${\cal F}_{11} (x_i, y_i, \mu)$  
is related to the functions~$J_{11} ({\cal P}, k_i, p_i)$ as 
follows: 
\begin{equation} 
J_{11} ({\cal P}, k_i, p_i) = \frac{M^{14}}{128} \, 
{\cal F}_{11} (x_i, y_i, \mu) .   
\label{eq:J-F-connection } 
\end{equation} 
The first integral in Eq.~(\ref{eq:dn11}) serves for a selection 
of the generated Monte Carlo events with the $\eta^\prime$-meson 
energy being inside the bin $z_i - \Delta z/2 < z < z_i + \Delta z/2$. 

In the same way all the other contributions from the matrix 
element squared were implemented in the numerical analysis.   
In particular, the contribution to the energy spectrum coming 
from the product $(1/3) \sum {\cal M}_1^* {\cal M}_2 +$ c.c. can 
be written as:
\begin{eqnarray}
\frac{\Delta n^{(12)}_i}{\Delta z} & = & 
\hspace*{-4mm} 
\int\limits_{z_i - \Delta z/2}^{z_i + \Delta z/2} \!\! dz \, 
\delta (z - 2 E_{\eta^\prime} / M)  
\int \frac{d{\bf k}_1}{2 \omega_1} \, \frac{d{\bf k}_2}{2 \omega_2} \,
\frac{d{\bf k}_3}{2 \omega_3} \, 
\frac{d{\bf p_{\eta^\prime}}}{2 E_{\eta^\prime}} \,
\delta^{(4)} ({\cal P} - k_1 - k_2 - k_3 - p_{\eta^\prime})  
\nonumber \\
& \times &
\frac{M^2 F_{\eta^\prime g} (y_1) F_{\eta^\prime g} (y_2)}
     {64 \pi^5 (\pi^2 - 9) M^4} \, 
\frac{2 {\cal F}_{12} (x_i, y_i, \mu)}
     {y_1 y_2 x_1 x_2 x_3^2 (2 - 2 y_1 -  x_2 - x_3)  
      (2 - 2 y_2 -  x_1 - x_3)} . 
\qquad   
\label{eq:dn12}  
\end{eqnarray} 
However, we have not been able to find compact expressions 
for the non-diagonal terms $(1/3) \sum {\cal M}_i {\cal M}_j^*$ 
($i \neq j$). We  present here instead the dimensionless 
function~${\cal F}_{12}(x_i,y_i,\mu)$ in a form of a series 
in powers of~$z_0^2/4=m_{\eta^\prime}^2/M^2$:
\begin{equation} 
{\cal F}_{12}(x_i, y_i, z_0) = 
\sum_{k=0}^5 f_{12}^k (x_i, y_i) \, 
\left ( \frac{z_0^2}{4} \right )^k, 
\label{eq:F12-decomposition} 
\end{equation} 
where functions~$f_{12}^k (x_i, y_i)$ are as follows: 

\begin{eqnarray}
 f_{12}^0 & = & \frac{1}{4} \bigg \{ 
(\Delta x_{12})^4 (\Delta y_{12})^2 - 4 (\Delta x_{12})^3 (\Delta y_{12})^3 + 4 (\Delta x_{12})^3 \Delta y_{12} y_3^2  
\label{eq:f12-0} \\ 
& + & 
  8 (\Delta x_{12})^3 \Delta y_{12} y_3 (x_{12} + x_3 - 2) 
\nonumber \\ 
& + & 
  4 (\Delta x_{12})^3 \Delta y_{12} (x_{12}^2 + 2 x_{12} x_3 - 4 x_{12} 
  + x_3^2 - 4 x_3 + 4) 
\nonumber \\ 
& + & 
  5 (\Delta x_{12})^2 (\Delta y_{12})^4 
+ 2 (\Delta x_{12})^2 (\Delta y_{12})^2 y_3 ( - 2 x_{12} - 4 x_3 + 7) 
\nonumber \\ 
& + & 
  2 (\Delta x_{12})^2 (\Delta y_{12})^2 ( - x_{12}^2 - 6 x_{12} x_3 
+ 7 x_{12} - 2 x_3^2 + 9 x_3 - 8) 
\nonumber \\ 
& - & 
  2 \Delta x_{12} (\Delta y_{12})^5 
+ 4 \Delta x_{12} (\Delta y_{12})^3 y_3^2 ( - x_3 - 2) 
\nonumber \\ 
& + & 
  4 \Delta x_{12} (\Delta y_{12})^3 y_3 ( - x_{12} x_3 - 2 x_{12} + 2 x_3) 
+ 4 \Delta x_{12} (\Delta y_{12})^3 x_{12} ( - x_{12} + x_3 + 1) 
\nonumber \\ 
& + & 
  2 \Delta x_{12} \Delta y_{12} y_3^4 (2 x_3 + 5) 
+ 4 \Delta x_{12} \Delta y_{12} y_3^3 (3 x_{12} x_3 + 2 x_{12} + 2 x_3^2 
  - 12 x_3 + 12) 
\nonumber \\ 
& + & 
  4 \Delta x_{12} \Delta y_{12} y_3^2 (3 x_{12}^2 x_3 + 4 x_{12} x_3^2 
  - 17 x_{12} x_3 + 19 x_{12} + x_3^3 - 11 x_3^2 + 36 x_3 - 36) 
\nonumber \\ 
& + &
  4 \Delta x_{12} \Delta y_{12} y_3 (x_{12}^3 x_3 + 2 x_{12}^2 x_3^2 
  - 12 x_{12}^2 x_3 + 14 x_{12}^2 + x_{12} x_3^3 - 16 x_{12} x_3^2 
\nonumber \\ 
&&
  + 54 x_{12} x_3 - 52 x_{12} - 4 x_3^3 + 28 x_3^2 - 64 x_3 + 48) 
\nonumber \\ 
& + & 
  2 \Delta x_{12} \Delta y_{12} ( - x_{12}^4 - 6 x_{12}^3 x_3 + 14 x_{12}^3
  - 8 x_{12}^2 x_3^2 + 44 x_{12}^2 x_3 - 56 x_{12}^2 - 2 x_{12} x_3^3 
\nonumber \\ 
&& 
  + 30 x_{12} x_3^2 - 96 x_{12} x_3 + 88 x_{12} + x_3^4 - 24 x_3^2 
  + 64 x_3 - 48) 
\nonumber \\ 
& + & 
  4 (\Delta y_{12})^4 y_3^2 
+ 2 (\Delta y_{12})^4 y_3 (3 x_{12} - 1) 
\nonumber \\ 
& + & 
  (\Delta y_{12})^4 (3 x_{12}^2 + 2 x_{12} x_3 - 6 x_{12} + 2 x_3^2 
  - 6 x_3 + 4) 
\nonumber \\ 
& - & 
  8 (\Delta y_{12})^2 y_3^4 
+ 4 (\Delta y_{12})^2 y_3^3 (x_{12} x_3 - 4 x_{12} + 2 x_3^2 + 2 x_3 - 7) 
\nonumber \\ 
& + & 
  4 (\Delta y_{12})^2 y_3^2 (2 x_{12}^2 x_3 - 2 x_{12}^2 + 3 x_{12} x_3^2 
  - 4 x_{12} x_3 - 7 x_{12} - 9 x_3^2 - 3 x_3 + 20) 
\nonumber \\ 
& + & 
  2 (\Delta y_{12})^2 y_3 (2 x_{12}^3 x_3 - 2 x_{12}^3 + 2 x_{12}^2 x_3^2 
  - 10 x_{12}^2 x_3 - 7 x_{12}^2 - 20 x_{12} x_3^2 - 4 x_{12} x_3 
\nonumber \\ 
&& 
  + 48 x_{12} + 10 x_3^2 + 32 x_3 - 52) 
\nonumber \\ 
& + & 
  (\Delta y_{12})^2 (x_{12}^4 - 14 x_{12}^3 - 12 x_{12}^2 x_3^2 
  - 10 x_{12}^2 x_3 + 56 x_{12}^2 - 8 x_{12} x_3^3 + 36 x_{12} x_3^2 
\nonumber \\ 
&& 
  + 32 x_{12} x_3 - 88 x_{12} - 4 x_3^4 + 20 x_3^3 - 32 x_3^2 - 24 x_3 + 48)
\nonumber \\ 
& - & 
  (\Delta x_{12})^4 y_3^2 
+ 2 (\Delta x_{12})^4 y_3 ( - x_{12} - x_3 + 2) 
\nonumber \\ 
& + & 
  (\Delta x_{12})^4 ( - x_{12}^2 - 2 x_{12} x_3 + 4 x_{12} - x_3^2 
  + 4 x_3 - 4) 
\nonumber \\ 
& + & 
  7 (\Delta x_{12})^2 y_3^4 
+ 2 (\Delta x_{12})^2 y_3^3 (8 x_{12} + 10 x_3 - 19) 
\nonumber \\ 
& + & 
  2 (\Delta x_{12})^2 y_3^2 (6 x_{12}^2 + 20 x_{12} x_3 - 33 x_{12} 
  + 11 x_3^2 - 43 x_3 + 40) 
\nonumber \\ 
& + & 
  2 (\Delta x_{12})^2 y_3 (2 x_{12}^3 + 16 x_{12}^2 x_3 - 21 x_{12}^2 
  + 20 x_{12} x_3^2 - 70 x_{12} x_3 + 56 x_{12} + 6 x_3^3 
\nonumber \\ 
&&
  - 37 x_3^2 + 72 x_3 - 44) 
\nonumber \\ 
& + & 
  \Delta x_{12}^2 (x_{12}^4 + 12 x_{12}^3 x_3 - 14 x_{12}^3 
  + 24 x_{12}^2 x_3^2 - 78 x_{12}^2 x_3 + 56 x_{12}^2 + 16 x_{12} x_3^3 
\nonumber \\ 
&&
  - 90 x_{12} x_3^2 + 160 x_{12} x_3 - 88 x_{12} + 3 x_3^4 - 26 x_3^3 
  + 80 x_3^2 - 104 x_3 + 48) 
\nonumber \\ 
& + &  
  4 y_3^6 + 2 y_3^5 ( - 2 x_{12} x_3 + 5 x_{12} - 4 x_3^2 - 4 x_3 + 15) 
\nonumber \\ 
& + & 
  y_3^4 ( - 16 x_{12}^2 x_3 + 9 x_{12}^2 - 36 x_{12} x_3^2 
  + 38 x_{12} x_3 + 50 x_{12} - 16 x_3^3 + 82 x_3^2 - 14 x_3 - 84) 
\nonumber \\ 
& + & 
  2 y_3^3 ( - 12 x_{12}^3 x_3 + 2 x_{12}^3 - 30 x_{12}^2 x_3^2 
  + 56 x_{12}^2 x_3 + 15 x_{12}^2 - 22 x_{12} x_3^3 + 126 x_{12} x_3^2 
\nonumber \\ 
&& 
  - 68 x_{12} x_3 - 64 x_{12} - 4 x_3^4 + 52 x_3^3 - 126 x_3^2 + 24 x_3 + 52) 
\nonumber \\ 
& + & 
  y_3^2 ( - 16 x_{12}^4 x_3 + x_{12}^4 - 44 x_{12}^3 x_3^2 
  + 104 x_{12}^3 x_3 + 10 x_{12}^3 - 40 x_{12}^2 x_3^3 + 298 x_{12}^2 x_3^2 
\nonumber \\ 
&&
  - 250 x_{12}^2 x_3 - 64 x_{12}^2 - 12 x_{12} x_3^4 + 228 x_{12} x_3^3 
  - 644 x_{12} x_3^2 + 288 x_{12} x_3 + 104 x_{12} 
\nonumber \\ 
&& 
  + 40 x_3^4 - 260 x_3^3 + 456 x_3^2 - 152 x_3 - 48) 
\nonumber \\ 
& + & 
  2 y_3 x_3 ( - 2 x_{12}^5 - 6 x_{12}^4 x_3 + 19 x_{12}^4 - 6 x_{12}^3 x_3^2 
  + 66 x_{12}^3 x_3 - 74 x_{12}^3 - 2 x_{12}^2 x_3^3 
\nonumber \\ 
&& 
  + 68 x_{12}^2 x_3^2 - 227 x_{12}^2 x_3 + 144 x_{12}^2 + 21 x_{12} x_3^3 
  - 156 x_{12} x_3^2 + 304 x_{12} x_3 
\nonumber \\ 
&&
  - 136 x_{12} - 17 x_3^3 + 88 x_3^2 - 132 x_3 + 48) 
\nonumber \\ 
& + & 
  x_3^2 (7 x_{12}^4 + 16 x_{12}^3 x_3 - 46 x_{12}^3 + 13 x_{12}^2 x_3^2 
  - 78 x_{12}^2 x_3 + 112 x_{12}^2 + 6 x_{12} x_3^3 
\nonumber \\ 
&& 
  - 46 x_{12} x_3^2 + 128 x_{12} x_3 - 120 x_{12} + 2 x_3^4 
  - 14 x_3^3 + 44 x_3^2 - 72 x_3 + 48)
\bigg \} , 
\nonumber 
\end{eqnarray} 
\begin{eqnarray}
 f_{12}^1 & = & \frac{1}{2} \bigg \{ 
- 8 (\Delta x_{12})^3 \Delta y_{12} y_3 
+ 4 (\Delta x_{12})^3 \Delta y_{12} ( - x_{12} - x_3 + 2) 
\label{eq:f12-1} \\ 
& + & 2 (\Delta x_{12})^2 (\Delta y_{12})^2 y_3 (x_3 + 7) 
+ (\Delta x_{12})^2 (\Delta y_{12})^2 (2 x_{12} x_3 + 4 x_{12} - 2 x_3 - 9) 
\nonumber \\ 
& - & 
  6 \Delta x_{12} (\Delta y_{12})^3 y_3 
+ 2 \Delta x_{12} (\Delta y_{12})^3 (x_3 + 2) 
- 2 \Delta x_{12} \Delta y_{12} y_3^3 (2 x_3 + 3) 
\nonumber \\
& + & 
  2 \Delta x_{12} \Delta y_{12} y_3^2 ( - 4 x_{12} x_3 - 2 x_{12} 
  - 2 x_3^2 + 7 x_3 - 10) 
\nonumber \\
& + & 
  2 \Delta x_{12} \Delta y_{12} y_3 ( - 2 x_{12}^2 x_3 - 3 x_{12}^2 
  - 2 x_{12} x_3^2 + 2 x_{12} x_3 + 2 x_{12} - x_3^2 + 8 x_3) 
\nonumber \\
& + & 
  2 \Delta x_{12} \Delta y_{12} ( - 2 x_{12}^3 + x_{12}^2 x_3 
  + 8 x_{12}^2 + 4 x_{12} x_3^2 - 2 x_{12} x_3 - 12 x_{12} 
  + x_3^3 
\nonumber \\
&&
  - 2 x_3^2 - 4 x_3 + 8) 
\nonumber \\
& - & 
  (\Delta y_{12})^4 (x_{12} + 1) 
+ 8 (\Delta y_{12})^2 y_3^3 
+ 2 (\Delta y_{12})^2 y_3^2 ( - x_{12} x_3 + 9 x_{12} + 2 x_3 + 5) 
\nonumber \\
& + & 
  4 (\Delta y_{12})^2 y_3 ( - x_{12}^2 x_3 + 2 x_{12}^2 + 2 x_{12} x_3 
  - 2 x_{12} - 6 x_3) 
\nonumber \\
& + & 
  (\Delta y_{12})^2 ( - 2 x_{12}^3 x_3 + 4 x_{12}^3 + 6 x_{12}^2 x_3 
  - 15 x_{12}^2 - 6 x_{12} x_3^2 + 16 x_{12} - 4 x_3^3 
\nonumber \\
&&
  + 18 x_3^2 - 8 x_3 - 4)
\nonumber \\
& + & 
  (\Delta x_{12})^4 y_3 + (\Delta x_{12})^4 (x_{12} + x_3 - 2) 
+ 2 (\Delta x_{12})^2 y_3^3 ( - x_3 - 8) 
\nonumber \\
& + & 
  (\Delta x_{12})^2 y_3^2 ( - 6 x_{12} x_3 - 22 x_{12} - 4 x_3^2 
  + 4 x_3 + 27) 
\nonumber \\
& + & 
  2 (\Delta x_{12})^2 y_3 ( - 3 x_{12}^2 x_3 - 3 x_{12}^2 
  - 4 x_{12} x_3^2 + 4 x_{12} x_3 + 11 x_{12} - x_3^3 + 6 x_3^2 - 3 x_3 - 8) 
\nonumber \\
& + & 
  (\Delta x_{12})^2 ( - 2 x_{12}^3 x_3 - 4 x_{12}^2 x_3^2 + 10 x_{12}^2 x_3 
  - x_{12}^2 - 2 x_{12} x_3^3 + 18 x_{12} x_3^2 - 30 x_{12} x_3 
\nonumber \\
&&
  + 8 x_{12} + 8 x_3^3 - 33 x_3^2 + 40 x_3 - 12) 
\nonumber \\
& - & 
8 y_3^5 + y_3^4 (6 x_{12} x_3 - 21 x_{12} + 8 x_3^2 + 4 x_3 - 41)
\nonumber \\
& + & 
  2 y_3^3 (10 x_{12}^2 x_3 - 9 x_{12}^2 + 14 x_{12} x_3^2 - 24 x_{12} x_3 
  - 24 x_{12} + 4 x_3^3 - 24 x_3^2 + 12 x_3 + 44) 
\nonumber \\
& + & 
  y_3^2 (24 x_{12}^3 x_3 - 6 x_{12}^3 + 36 x_{12}^2 x_3^2 - 88 x_{12}^2 x_3 
  - 19 x_{12}^2 + 14 x_{12} x_3^3 - 94 x_{12} x_3^2 
\nonumber \\ 
&&
  + 56 x_{12} x_3 + 112 x_{12} - 16 x_3^3 + 50 x_3^2 + 32 x_3 - 100) 
\nonumber \\
& + & 
  y_3 (12 x_{12}^4 x_3 - x_{12}^4 + 20 x_{12}^3 x_3^2 - 56 x_{12}^3 x_3 
  - 10 x_{12}^3 + 8 x_{12}^2 x_3^3 - 86 x_{12}^2 x_3^2 + 82 x_{12}^2 x_3 
\nonumber \\
&&
  + 64 x_{12}^2 - 32 x_{12} x_3^3 + 152 x_{12} x_3^2 - 56 x_{12} x_3 
  - 104 x_{12} + 40 x_3^3 - 120 x_3^2 + 40 x_3 + 48) 
\nonumber \\
& + & 
  x_3 (2 x_{12}^5 + 4 x_{12}^4 x_3 - 19 x_{12}^4 + 2 x_{12}^3 x_3^2 
  - 34 x_{12}^3 x_3 + 74 x_{12}^3 - 8 x_{12}^2 x_3^2 + 97 x_{12}^2 x_3 
  - 144 x_{12}^2 
\nonumber \\
&& 
  + 11 x_{12} x_3^3 - 8 x_{12} x_3^2 - 104 x_{12} x_3 
  + 136 x_{12} + 4 x_3^4 - 25 x_3^3 + 32 x_3^2 + 28 x_3 - 48)
\bigg \} , 
\nonumber 
\end{eqnarray}
\begin{eqnarray}
f_{12}^2 & = & \frac{1}{4} \bigg \{ 
- 4 (\Delta x_{12})^3 \Delta y_{12} 
- 4 (\Delta x_{12})^2 (\Delta y_{12})^2 
+ 4 \Delta x_{12} (\Delta y_{12})^3 
\label{eq:f12-2} \\ 
& + &
  4 \Delta x_{12} \Delta y_{12} y_3^2 (x_3 - 2) 
+ 4 \Delta x_{12} \Delta y_{12} y_3 (x_{12} x_3 - 2 x_{12} - 6 x_3 + 8) 
\nonumber \\
& + & 
  4 \Delta x_{12} \Delta y_{12} (3 x_{12}^2 - x_{12} x_3 - x_{12} + 4 x_3 - 4) 
- 8 (\Delta y_{12})^2 y_3^2 
- 12 (\Delta y_{12})^2 y_3 (2 x_{12} + 1) 
\nonumber \\
& + & 
  4 (\Delta y_{12})^2 ( - 2 x_{12}^2 + x_{12} - x_3^2 + 5 x_3)
+ 3 (\Delta x_{12})^4 + 2 (\Delta x_{12})^2 y_3^2 (4 x_3 + 23) 
\nonumber \\
& + & 
  2 (\Delta x_{12})^2 y_3 (8 x_{12} x_3 + 20 x_{12} + 4 x_3^2 - 6 x_3 - 17) 
\nonumber \\
& + & 
  2 (\Delta x_{12})^2 (4 x_{12}^2 x_3 + 4 x_{12} x_3^2 + 4 x_{12} x_3 
  - 9 x_{12} + 5 x_3^2 - 27 x_3 + 16) 
\nonumber \\
& + &
  24 y_3^4 + 4 y_3^3 ( - 3 x_{12} x_3 + 17 x_{12} - 2 x_3^2 + 2 x_3 + 21) 
\nonumber \\
& + & 
  2 y_3^2 ( - 16 x_{12}^2 x_3 + 27 x_{12}^2 - 10 x_{12} x_3^2 
  + 46 x_{12} x_3 + 22 x_{12} + 22 x_3^2 - 50 x_3 - 52) 
\nonumber \\
& + & 
  2 y_3 ( - 14 x_{12}^3 x_3 + 6 x_{12}^3 - 10 x_{12}^2 x_3^2 
  + 44 x_{12}^2 x_3 - 7 x_{12}^2 + 18 x_{12} x_3^2 - 12 x_{12} x_3 
\nonumber \\
&& 
  - 32 x_{12} + 10 x_3^2 - 56 x_3 + 44) 
\nonumber \\
& - & 
  8 x_{12}^4 x_3 + x_{12}^4 - 8 x_{12}^3 x_3^2 + 8 x_{12}^3 x_3 
+ 10 x_{12}^3 + 18 x_{12}^2 x_3^2 + 86 x_{12}^2 x_3 - 64 x_{12}^2 
\nonumber \\
& + & 
  28 x_{12} x_3^3 - 52 x_{12} x_3^2 - 176 x_{12} x_3 + 104 x_{12} 
+ 12 x_3^4 - 68 x_3^3 + 88 x_3^2 + 72 x_3 - 48
\bigg \} , 
\nonumber 
\end{eqnarray}
\begin{eqnarray}
 f_{12}^3 & = & \frac{1}{2} \bigg \{ 
  2 (\Delta y_{12})^2 (x_{12} + 1) 
+ 2 \Delta y_{12} \Delta x_{12} (3 y_3 + 2 x_{12} - x_3 - 2) 
\label{eq:f12-3} \\ 
& - & 
  2 (\Delta x_{12})^2 y_3 (x_3 + 6)  
  + (\Delta x_{12})^2 (- 2 x_{12} x_3 - 6 x_{12} + 8 x_3 + 1)
\nonumber \\
& - &
  8 y_3^3 + 2 y_3^2 (x_{12} x_3 - 13 x_{12} - 2 x_3 - 11) 
  + 2 y_3 (2 x_{12}^2 x_3 - 9 x_{12}^2 - 8 x_{12} x_3 + 20 x_3 + 4) 
\nonumber \\
& + & 
    2 x_{12}^3 x_3 - 2 x_{12}^3  
 - 12 x_{12}^2 x_3 + 11 x_{12}^2 + 6 x_{12} x_3^2  
 + 24 x_{12} x_3 - 16 x_{12} + 4 x_3^3 - 22 x_3^2 + 4
\bigg \} , 
\nonumber \\ 
 f_{12}^4 & = & \frac{1}{4} \bigg \{
 3 (\Delta x_{12})^2 - 2 \Delta y_{12} \Delta x_{12} 
+ 4 y_3^2 + 2 y_3 (9 x_{12} + 7) 
\label{eq:f12-4} \\
& + &
  9 x_{12}^2 - 2 x_{12} x_3 + 2 x_{12} + 2 x_3^2 - 14 x_3 - 4 
\bigg \} , 
 \nonumber \\ 
 f_{12}^5 & = & - \frac{1}{2} ( x_{12} + 1 ) , 
\label{eq:f12-5} 
\end{eqnarray}
where $x_{12} = x_1 + x_2$, $\Delta x_{12} = x_1 - x_2$, and 
$\Delta y_{12} = y_1 - y_2$. It is easy to see that the equations 
for $f_{12}^k$ are symmetric under the interchange 
$(x_1, y_1) \leftrightarrow (x_2, y_2)$.  

The expressions for the other~${\cal F}_{ij}(x_i,y_i,\mu)$ can be 
derived from the ones given above by using the Bose symmetry.


\begin{thebibliography}{22} 
\bibitem{Gilman:1987ax}
F.~J.~Gilman and R.~Kauffman,
Phys.\ Rev.\ D {\bf 36}, 2761 (1987)   
[Erratum-ibid.\ D {\bf 37}, 3348 (1988)].

\bibitem{Leutwyler:1997yr}
H.~Leutwyler,
Nucl.\ Phys.\ Proc.\ Suppl.\  {\bf 64}, 223 (1998)
[arXiv:hep-ph/9709408].

\bibitem{Kaiser:2000gs}
R.~Kaiser and H.~Leutwyler,
Eur.\ Phys.\ J.\ C {\bf 17}, 623 (2000)
[arXiv:hep-ph/0007101].

\bibitem{Feldmann:1998vh}
T.~Feldmann, P.~Kroll and B.~Stech,
Phys.\ Rev.\ D {\bf 58}, 114006 (1998)
[arXiv:hep-ph/9802409].

\bibitem{Feldmann:1998sh}
T.~Feldmann, P.~Kroll and B.~Stech,
Phys.\ Lett.\ B {\bf 449}, 339 (1999)
[arXiv:hep-ph/9812269].

\bibitem{Ball:1995zv}
P.~Ball, J.~M.~Frere and M.~Tytgat,
Phys.\ Lett.\ B {\bf 365}, 367 (1996)
[arXiv:hep-ph/9508359].

\bibitem{Atwood:1997bn}
D.~Atwood and A.~Soni,
Phys.\ Lett.\ B {\bf 405}, 150 (1997)
[arXiv:hep-ph/9704357].

\bibitem{Hou:1997wy}
W.~S.~Hou and B.~Tseng,
Phys.\ Rev.\ Lett.\  {\bf 80}, 434 (1998) 
[arXiv:hep-ph/9705304].

\bibitem{Kagan:1997qn}
A.~L.~Kagan and A.~A.~Petrov,
arXiv:hep-ph/9707354.

\bibitem{Halperin:1997ma}
I.~E.~Halperin and A.~Zhitnitsky,
Phys.\ Rev.\ Lett.\  {\bf 80}, 438 (1998)
[arXiv:hep-ph/9705251].

\bibitem{Yuan:1997ts}
F.~Yuan and K.~T.~Chao,
Phys.\ Rev.\ D {\bf 56}, 2495 (1997)
[arXiv:hep-ph/9706294].

\bibitem{Datta:1997nr}
A.~Datta, X.~G.~He and S.~Pakvasa,
Phys.\ Lett.\ B {\bf 419}, 369 (1998)
[arXiv:hep-ph/9707259].

\bibitem{Ahmady:1997fa}
M.~R.~Ahmady, E.~Kou and A.~Sugamoto,   
Phys.\ Rev.\ D {\bf 58}, 014015 (1998)
[arXiv:hep-ph/9710509].

\bibitem{Halperin:1997as}
I.~E.~Halperin and A.~Zhitnitsky,
Phys.\ Rev.\ D {\bf 56}, 7247 (1997)
[arXiv:hep-ph/9704412].

\bibitem{Cheng:1997if}
H.~Y.~Cheng and B.~Tseng,
Phys.\ Lett.\ B {\bf 415}, 263 (1997)
[arXiv:hep-ph/9707316].

\bibitem{Dighe:1997hm}
A.~S.~Dighe, M.~Gronau and J.~L.~Rosner,
Phys.\ Rev.\ Lett.\  {\bf 79}, 4333 (1997)
[arXiv:hep-ph/9707521].

\bibitem{Deshpande:1997ar}
N.~G.~Deshpande, B.~Dutta and S.~Oh,
Phys.\ Rev.\ D {\bf 57}, 5723 (1998)
[arXiv:hep-ph/9710354].

\bibitem{Du:1997hs}
D.~s.~Du, C.~S.~Kim and Y.~d.~Yang,
Phys.\ Lett.\ B {\bf 426}, 133 (1998)
[arXiv:hep-ph/9711428].

\bibitem{Ali:1997ex}
A.~Ali, J.~Chay, C.~Greub and P.~Ko,
Phys.\ Lett.\ B {\bf 424}, 161 (1998)
[arXiv:hep-ph/9712372].

\bibitem{Ali:1998eb}
A.~Ali, G.~Kramer and C.~D.~Lu,
Phys.\ Rev.\ D {\bf 58}, 094009 (1998)
[arXiv:hep-ph/9804363].

\bibitem{Beneke:2002jn}
M.~Beneke and M.~Neubert,
Nucl.\ Phys.\ B {\bf 651}, 225 (2003)
[arXiv:hep-ph/0210085].

\bibitem{Kagan:2002dq}
A.~L.~Kagan,
AIP Conf.\ Proc.\  {\bf 618}, 310 (2002)
[arXiv:hep-ph/0201313]; Y. Chen and A.L. Kagan, Univ. of Cincinnati 
preprint
(in preparation).


\bibitem{Muta:1999tc}
T.~Muta and M.~Z.~Yang,
Phys.\ Rev.\ D {\bf 61}, 054007 (2000)  
[arXiv:hep-ph/9909484].

\bibitem{Ali:2000ci}
A.~Ali and A.~Y.~Parkhomenko,
Phys.\ Rev.\ D {\bf 65}, 074020 (2002)
[arXiv:hep-ph/0012212].

\bibitem{Kroll:2002nt}
P.~Kroll and K.~Passek-Kumericki,
Phys.\ Rev.\ D {\bf 67}, 054017 (2003)
[arXiv:hep-ph/0210045].

\bibitem{Agaev:2002ek} 
S.~S.~Agaev and N.~G.~Stefanis,
arXiv:hep-ph/0212318. 

\bibitem{Terentev:qu}
M.~V.~Terentev,
Sov.\ J.\ Nucl.\ Phys.\  {\bf 33}, 911 (1981)
[Yad.\ Fiz.\  {\bf 33}, 1692 (1981)].

\bibitem{Ohrndorf:1981uz}
T.~Ohrndorf,
Nucl.\ Phys.\ B {\bf 186}, 153 (1981).

\bibitem{Shifman:1981dk} 
M.A.~Shifman and M.I.~Vysotsky,
Nucl.\ Phys.\ B {\bf 186}, 475 (1981).

\bibitem{Baier:1981pm}
V.N.~Baier and A.G.~Grozin,
Nucl.\ Phys.\ B {\bf 192}, 476 (1981).

\bibitem{Terentev:wv}
M.~V.~Terentev,
JETP\ Lett.\ {\bf 33}, 67 (1981)
[Pisma Zh.\ Eksp.\ Teor.\ Fiz.\  {\bf 33}, 71 (1981)].

\bibitem{Belitsky:1998gc}
A.~V.~Belitsky and D.~Muller,
Nucl.\ Phys.\ B {\bf 537}, 397 (1999)
[arXiv:hep-ph/9804379].

\bibitem{Gronberg:1997fj}
J.~Gronberg {\it et al.}  [CLEO Collaboration],
Phys.\ Rev.\ D {\bf 57}, 33 (1998)
[arXiv:hep-ex/9707031].

\bibitem{Acciarri:1997yx}
M.~Acciarri {\it et al.}  [L3 Collaboration],
Phys.\ Lett.\ B {\bf 418}, 399 (1998).

\bibitem{Artuso:2002px}
M.~Artuso {\it et al.}  [CLEO Collaboration],
Phys.\ Rev.\ D {\bf 67}, 052003 (2003)
[arXiv:hep-ex/0211029].

\bibitem{Mackenzie:1981sf}
P.~B.~Mackenzie and G.~P.~Lepage,
Phys.\ Rev.\ Lett.\  {\bf 47}, 1244 (1981).

\bibitem{Feldmann:1999uf}
T.~Feldmann,
Int.\ J.\ Mod.\ Phys.\ A {\bf 15}, 159 (2000)
[arXiv:hep-ph/9907491].

\bibitem{Feldmann:2002kz}
T.~Feldmann and P.~Kroll,
Phys.\ Scripta {\bf T99}, 13 (2002)
[arXiv:hep-ph/0201044].

\bibitem{Ali:2003}   
A.~Ali and A.~Y.~Parkhomenko,
``The $\eta^\prime g^* g^*$ Vertex 
Including the $\eta^\prime$-Meson Mass'', 
CERN-TH/2003-063 (to be published).

\bibitem{James:1975dr}
F.~James and M.~Roos,
Comput.\ Phys.\ Commun.\  {\bf 10} (1975) 343.


\end{thebibliography}
\end{document}